\documentclass[10pt,journal,compsoc]{IEEEtran}
\IEEEoverridecommandlockouts

\usepackage{cite}
\usepackage{amsmath,amssymb,amsfonts}
\usepackage{algorithmic}
\usepackage{graphicx}
\usepackage{textcomp}
\usepackage{xcolor}
\usepackage{xspace}
\usepackage{array,colortbl,multirow,multicol,booktabs,ctable}

\makeatletter
  \newcommand\tinyv{\@setfontsize\tinyv{7.2pt}{6}}
  \newcommand\tinycode{\@setfontsize\tinycode{8.2pt}{6}}
\makeatother

\newcommand{\ourwork}{\texttt{ServiceAudit}\xspace}

\def\BibTeX{{\rm B\kern-.05em{\sc i\kern-.025em b}\kern-.08em
    T\kern-.1667em\lower.7ex\hbox{E}\kern-.125emX}}
\begin{document}

\title{A Systematic Study of Android Non-SDK (Hidden) Service API Security}

\author{Yi He, Yacong~Gu, ~Purui~Su,  Kun Sun,~\IEEEmembership{Senior~Member,~IEEE,} Yajin Zhou,~\IEEEmembership{Member,~IEEE,}\\Zhi Wang,~\IEEEmembership{Member,~IEEE,} Qi~Li,~\IEEEmembership{Senior Member,~IEEE}

\thanks{H. Yi and Q. Li are with Institute for Network Sciences and Cyberspace and Beijing National Research Center for Information Science and Technology, Tsinghua University, Beijing, e-mail: \{heyi21@mails., qli01\}@tsinghua.edu.cn}
\thanks{Y. Gu and P. Su are with Institute of Software, Chinese Academy of Sciences,  China 100190, e-mail: \{guyangcong, supurui\}@tca.iscas.ac.cn.}
\thanks{K. Sun is with Center for Secure Information Systems, George Mason University, 4400 University Drive, Fairfax, VA 22030-4422, e-mail: ksun3@gmu.edu.}
\thanks{Y. Zhou is with Institute of Cyber Security Research, Zhejiang University, China 310058, email: zhouyajin@gmail.com.}
\thanks{Z. Wang is with Department of Computer Science, Florida State University, FL 32306, email: zwang@cs.fsu.edu.}
\thanks{This paper is extended from the previous conference paper~\cite{Gu16}. }
}

\IEEEcompsoctitleabstractindextext{
\begin{abstract}

Android allows apps to communicate with its system services via system service helpers so that these apps can use various functions provided by the system services. Meanwhile, the system services rely on their service helpers to enforce security checks for protection.  Unfortunately, the security checks in the service helpers may be bypassed via directly exploiting the non-SDK (hidden) APIs, degrading the stability and posing severe security threats such as privilege escalation, automatic function execution without users' interactions, crashes, and DoS attacks. Google has proposed various approaches to address this problem, e.g., case-by-case fixing the bugs or even proposing a blacklist to block all the non-SDK APIs. However, the developers can still figure out new  ways of exploiting these hidden APIs to evade the non-SDKs restrictions.  
In this paper, we systematically study the vulnerabilities due to the hidden API exploitation and analyze the effectiveness of Google’s countermeasures. We aim to answer if there are still vulnerable hidden APIs that can be exploited in newest Android 12. We develop a static analysis tool called \ourwork to automatically mine the inconsistent security enforcement between service helper classes and the hidden service APIs. We apply \ourwork to Android 6$\sim$12. Our tool discovers 112 vulnerabilities in Android 6 with a higher precision than existing approaches. Moreover, in Android 11 and 12, we identify more than 25 hidden APIs with inconsistent protections; however, only one of the vulnerable APIs can lead to severe security problem in Android 11, and none of them work on Android 12. 




\end{abstract}

}

\maketitle


\def\red#1{\textcolor{red}{#1}}

\section{Introduction}
%

\begin{figure}[ht]
\vspace{-10px}
\centering
\includegraphics[width=0.46\textwidth]{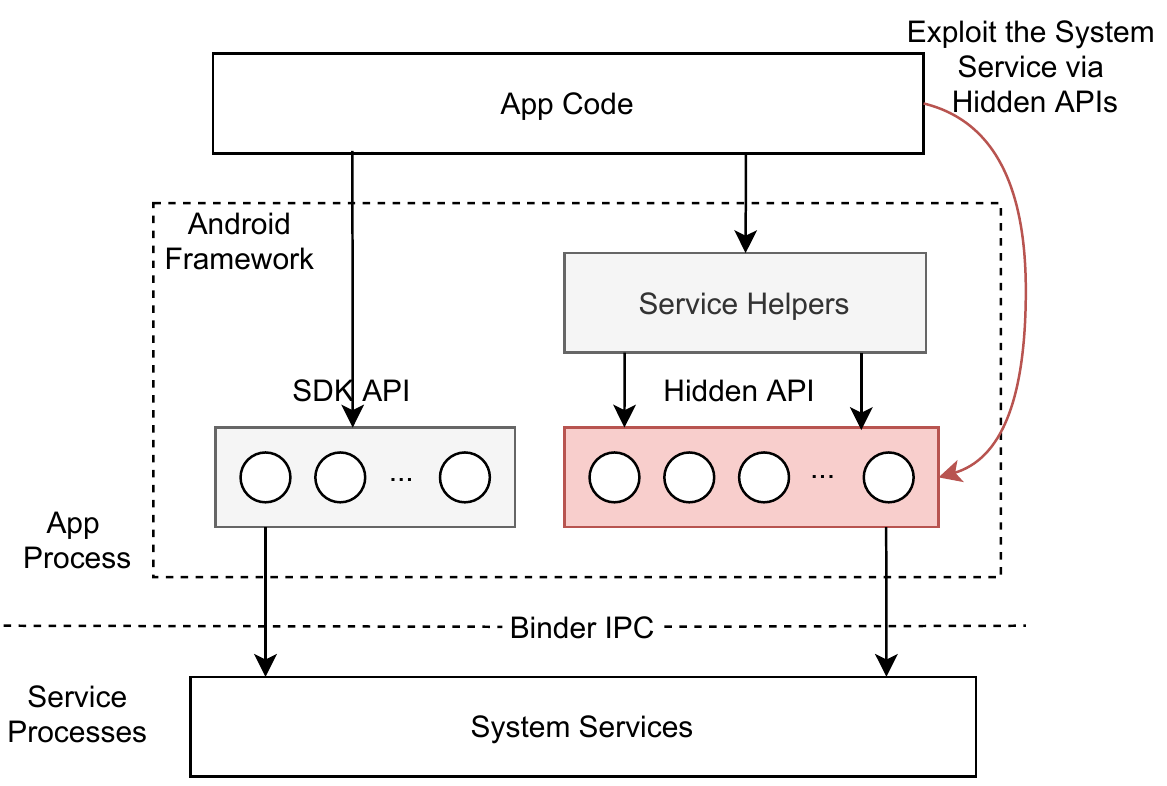}
\caption{Developers can bypass the service helpers by exploiting the hidden APIs to access system services.}
\label{fig:framework}
\end{figure}

Android has a modular binder-based inter-process communication (IPC) architecture that allows third-party apps to access many important functionalities of Android system services, e.g., telephony, notification, and clipboard.  Normally, apps can access these system services via the service helper classes which encapsulate the origin service IPC APIs which are hidden to the developers. As shown in Fig.~\ref{fig:framework}, these service helpers are loaded into the app's address space as a runtime library (\textit{framework.jar}) and developers are required to only use the public APIs exposed by the Android SDK (\textit{android.jar}) to build their apps. However, previous work~\cite{hidden_api_stat} shows that it is prevalent for the developers to directly exploit the non-SDK (hidden) APIs (includes internal and private APIs) in \textit{framework.jar} via Java reflection for the requirements of developing or hacking. This may lead to serious stability problems~\cite{improve_stability, hidden_api_stat} when the hidden APIs are changed in new Android versions but these Apps do not adapt to the new APIs in time, as well as security problems due to service helpers bypassing, especially when these residual APIs are not well maintained and have inconsistent security enforcements~\cite{Gu16, acedroid, residual_apis}. Since Android 9, for the sake of stability and security, Google has restricted~\cite{block_hidden_api} the usages of the hidden APIs by proposing a graylist includes the APIs which are available now but will be disabled in future versions, and a blacklist includes the APIs which are blocked now. Unfortunately, this restriction have been broken by the developers as soon as it appears. Even in the newest Android 12, developers have proposed several methods to evade the hidden APIs restriction and there are still some apps~\cite{bypass_blacklist} try to exploit various blacklist hidden APIs. 

The hidden API restriction is significant to Android's app development ecosystem. Since the Android framework APIs change dramatically in every version, the hidden APIs are also involved. Only a few of the hidden APIs can be promoted to permanent public APIs~\cite{hidden_api_stat} and most of them are removed in the newer versions. The hidden APIs are not stable to the developers and building apps on these temporary hidden APIs can lead to serious compatibility issues. Moreover, some hidden APIs without proper security enforcement may be abused by malicious apps. Thus, Google discourages the developers to directly exploit the hidden APIs~\cite{improve_stability} and propose a blacklist to block these APIs.  In Android 12, there are about 495 thousand APIs in the \textit{framework.jar} with only 12\% of them are public to the developers provided in the Android SDK (\textit{android.jar}), and over 249 thousand APIs (50\%) are blocked by the hidden APIs restriction policy. The hidden APIs can be used to access system resources and some of them can invoke system service via IPC. For the hidden APIs which invoke service IPC interfaces, it is quite common for them to lack security checks e.g. permission checking, input validation, and caller authentication, and lead to inconsistent security enforcement. Existing works, such as Kratos~\cite{ss_attack_2} and AceDroid~\cite{acedroid}, have discovered these kinds of problems in the public APIs of multiple Android SDK versions. Our previous work~\cite{Gu16} indicates that this problem also widely exists in the hidden APIs which invokes service interfaces via IPC calls. Google has taken measures to mitigate this problem since 2017 and put most of the hidden APIs into blacklist since Android 9.  Nevertheless, taking the huge number of hidden APIs, it is extremely difficult for Google to eliminate this problem by fixing all the buggy hidden APIs. In the newest Android 12, developers can still exploit the hidden APIs after bypassing the blacklist policy. Considering these developers can actually use all the APIs in the \textit{framework.jar} rather than just use the public API in android SDK, therefore, the vulnerable hidden APIs may still be exploitable.

In this paper, we focus on the non-SDK service API security and the effectiveness of the hidden API blacklist policy in newest Android versions (e.g., Android 10 $\sim$ 12). The hazards of exploiting hidden APIs has been also studied in previous research~\cite{Gu16, invetter, residual_apis}. Zeinab et al.~\cite{residual_apis} study the access control vulnerabilities in residual APIs that are erratically added or removed among different Android SDK versions.   Our previous work~\cite{Gu16} focuses on discovering inconsistent security enforcement between service helper classes and the service implementations. We found that exploiting hidden APIs can bypass all security enforcement in service helpers and the attackers can perform various attacks, e.g., bypassing the access control checks, sending fake identities or parameters to exploit vulnerable system services, or sending duplicate requests to perform DoS attacks. Invetter~\cite{invetter} addresses a similar issue. However, it focuses on detecting only a narrow type of input validation. Since during the evolution of Android framework, some of the vulnerable hidden APIs may be removed or mitigated by Google via enforcing the security checks. We seek to answer whether the newest Android has eventually eliminated the exploitable inconsistent security enforcement vulnerabilities in hidden service APIs. 

To achieve this goal, we perform a thorough security analyze on Android 10, 11, and 12, to detect if there are inconsistent security enforcement in hidden APIs and if these APIs are blocked by blacklist. We propose a new static analysis tool \texttt{ServiceAudit} to automatically discover APIs without insufficient security enforcement. Compared to our previous approach~\cite{Gu16}, we have adopted new analysis approaches to discover more security enforcements and significantly improved the precision. Specifically, to keep up with the changes of the Android framework, we modified \ourwork to support newest Android 12 and tracked if vulnerable hidden APIs in an earlier Android version can still be exploitable in newer versions. 


\textbf{Our Findings. } In Android 6, \ourwork has dicovered 112 vulnerabilities with more than 32 exploitable ones. In Android 11 and 12, less than 27 vulnerabilities are reported and most of them seem to have no harm. In particular, only one of them can incur security problems on Android 11 and are immediately fixed by Android 12. To understand why the vulnerabilities are eliminated, we perform case studies to check if the former vulnerable APIs can still be attacked. We find that Google has successfully addressed almost all of them by manually adding extra checks to system service. We further study if the blacklist can make the hidden APIs more secure and find that the attackers have no incentive to exploit these APIs by evading the hidden APIs restriction. 




\section{Background}
\label{sec:background}
Non-SDK APIs are hidden to developers and only accessible to the system code. However, developers can easily utilize these methods. Since Android 9, Google has proposed a blacklist policy to gradually limit the usages of these APIs and prevent the knowing hidden API exploiting approaches. In this section, we first introduce the concept of the hidden APIs and service helpers. Then we illustrate Google's hidden API restriction policy.

\begin{figure*}[ht!]
\centering
\includegraphics[width=0.8\textwidth]{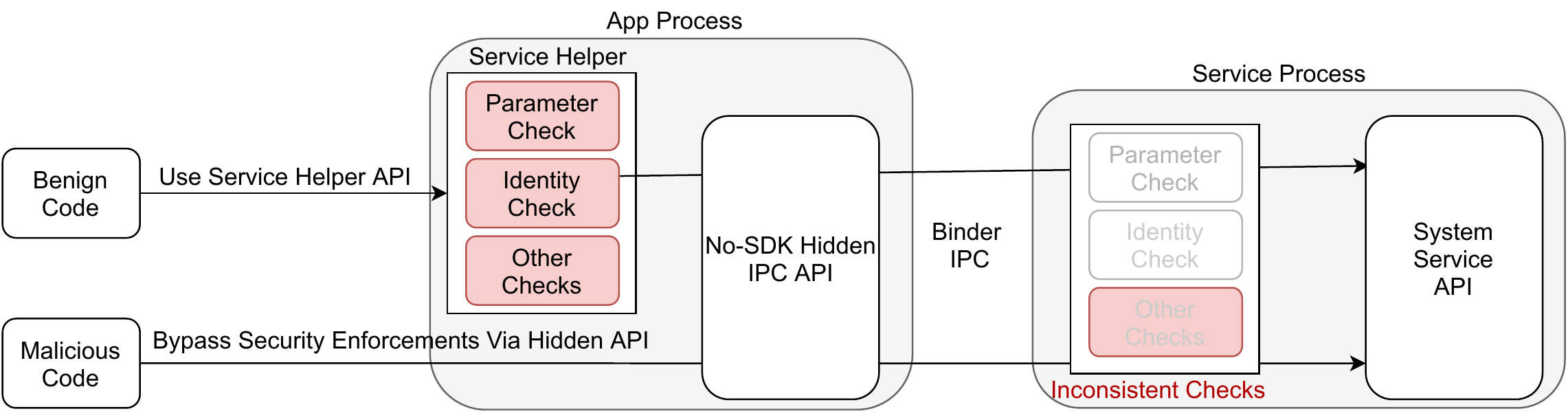}
\caption{Attackers can invoke the hidden APIs to bypass the security enforcements and sanity checks for the arguments in service helper APIs.  }
\vspace{-10px}
\label{fig:attack}
\end{figure*}

\subsection{Android Hidden APIs and Service Helpers}
Android apps are built on Android SDK (provided by \textit{android.jar}) during development and run on the Android framework (provided by \textit{framework.jar}) on real devices. The Android SDK only contains the public APIs that are open to the developers while the non-SDK APIs, aka Hidden APIs, are the inaccessible~\cite{hidden_api_stat} methods of Android framework that do not expose to developers but are used by the framework itself or the system apps. Android adopts a Binder based IPC architecture and provides AIDL~\cite{aidl} files to automatically generate the IPC server methods (\texttt{IPC.stub}) and the IPC client methods (\texttt{IPC.proxy}). System services implement the IPC server methods and custom code need to invoke these methods via the IPC client code. As a part of the Android SDK, service helpers encapsulate the complex origin IPC client code and provide easy-to-use interfaces for apps to access the IPC methods of system services, e.g., automatically feed in function parameters for invoking the origin service APIs. To protect system services against attacks, the corresponding service helpers also implement some security mechanisms to validate the requests from apps before issuing an IPC call to the system services.



As shown in Fig~\ref{fig:framework}, normally, developers only need to use the SDK APIs and access the System services via service helpers. However, once an App is installed and executed, the Android framework is loaded to the same process as the App’ user code. Therefore the App can actually access all the framework APIs including non-SDK APIs in \textit{framework.jar} via Java reflection or using a custom Android SDK file that contain all the internal methods of the Android framework~\cite{full_sdk}. Supposing the system service side's security enforcement is not as robust as the service helper~\cite{invetter, Gu16}, as shown in Fig~\ref{fig:attack}, the attackers can directly use hidden APIs to exploit the system services without enough protection to bypass all the security enforcements in service helpers' side and perform their attacks, e.g., DoS, privilege escalation.



\subsection{Google's Hidden APIs Restriction Policy}
The non-SDK APIs have been used widely by the developers~\cite{hidden_api_stat}. Since these APIs are changed rapidly among different SDk versions, using these APIs may affect the stability of apps and bring in security risks to the Android system. Google does not expect developers to directly use them and proposes a blacklist policy to restrict the hidden APIs usages in third-party apps since Android 9~\cite{block_hidden_api}. It is impossible to block all the hidden APIs immediately, considering many of them are already used by a larger number of apps. Google divided all the hidden APIs into three parts, the \texttt{whitelist} APIs that can be normally used, and the \texttt{graylist} APIs that can be used now but will be disallowed in the future, and the \texttt{blacklist} APIs that should not be used. That is to say, some of the vulnerabilities may still be exploitable in the newest Android if they are in the whitelist or graylist. Moreover, the attacker can even evade the blacklist~\cite{bypass_blacklist} to invoke all the non-SDK hidden APIs (see Section~\ref{sec:evade}).

\section{Problem Statement}
The hazards of exploiting hidden service APIs have been reported to Google since 2017 and Google have propose several measures to address these issues. In this section, we first introduce the hazards and the exploitation of the vulnerable hidden APIs. Then we summarize Google's countermeasures and the changes of newest Android.

\subsection{The Hazards of Exploiting Hidden APIs}
\label{sec:hazards}
To ensure security, reliability, and efficiency of system services, both the service helpers and the system service APIs need to enforce various security mechanisms. However, some security mechanisms may only be implemented on the system service's side and the corresponding service API does not check again for these enforcements. We have identified various vulnerability types in service APIs due to inconsistent security enforcement. As shown in Fig~\ref{fig:attack}, attackers can exploit these vulnerabilities in various ways and lead to privilege escalation, function execution without users’ interactions, system service crash, or DoS attacks. To better understand these hazards, we use real examples from AOSP to illustrate the root causes and exploiting methods of these vulnerabilities. 

\subsubsection{Privilege Escalation and Privilege Leakage}
For some service API, the access control checks are only applied on the service helpers’ side. Once the attackers exploit the service API directly, they can bypass all the checks in service helpers and perform privileged operations without corresponding permissions. We also found some system services just use values (caller identity such as uid, pid, and package name) passed by the service helpers to perform permission checks. Attackers can forge these identities by invoking the hidden service API with fake parameters, e.g., using the system app’s uid. Thus, these services are cheated by the attackers to check the permission of the system app rather than the attackers’ app. The detailed attacks and the corresponding vulnerable code are as follows.

\begin{figure}[htbp]
\vspace{-10px}
\centering
\includegraphics[width=0.40\textwidth]{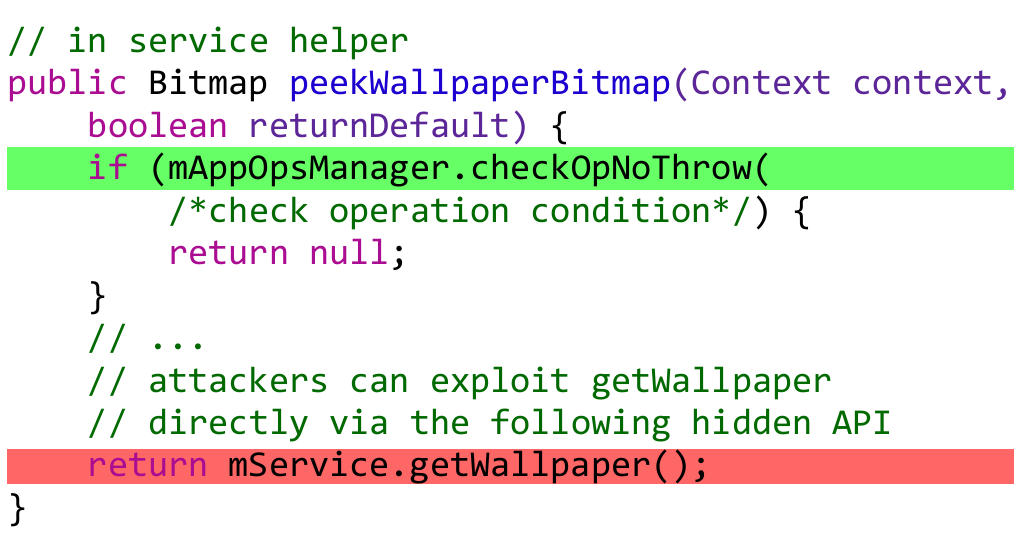}
\vspace{-10px}
\caption{Attackers can bypass the permission checks by directly invoke the hidden service API.}
\label{fig:code1}
\end{figure}

\noindent \textbf{Evading the Operation Permission Checks.} As shown in Fig~\ref{fig:code1}, the service helper checks the setup in \textit{WallpaperManager.peekWallpaperBitmap()} and tests if an app is authorized to access the wallpaper data. However, this security check is only performed in the service helper. A malicious app can bypass the helper and directly invoke the hidden API, i.e. \textit{WallpaperManagerService.getWallpaper()}. Thereby, the app can read the wallpaper data even if its access to the data is disabled by the system, which incurs private data leakage.

\begin{figure}[htbp]
\vspace{-15px}
\centering
\includegraphics[width=0.40\textwidth]{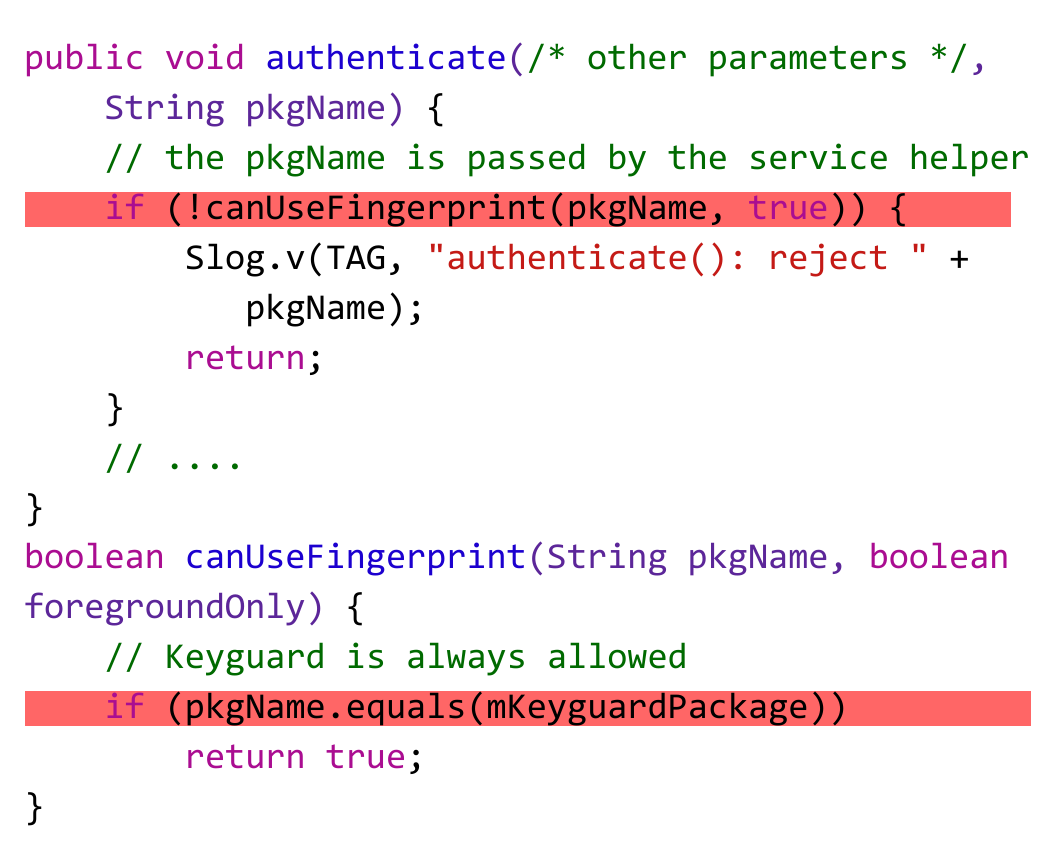}
\vspace{-10px}
\caption{Attackers can use fake package name to invoke the hidden service API.}
\label{fig:code-1-2}
\vspace{-10px}
\end{figure}

\noindent \textbf{Feeding Fake Identity to Service API.} Let us take the function of \textit{FingerprintService.authenticate()} as an example, the service helper class of \textit{FingerprintService}, i.e., \textit{FingerprintManager}, is responsible for automatically collecting and passing the caller's package name to 
\textit{FingerprintService.authenticate()} is used to authenticate a given fingerprint.
As shown in Fig~\ref{fig:code-1-2}, \textit{authenticate()}  verifies whether the caller is allowed to use fingerprint based on received package name. 
In \textit{canUseFingerprint()}, it evaluates whether the caller is the current user or specified by the current user profile, whether \textit{AppOps} allows the operation, and whether the caller is currently in the foreground.
However, if the caller's package name is of \textit{KeyguardService}, the caller will be always allowed to use the fingerprint and thus the above restrictions will be bypassed (the red line).
Unfortunately, \textit{authenticate()} does not verifies the authenticity of received package names. 

\subsubsection{Bypassing User Interaction}
Some Android system services may be allowed only when the callers, i.e., the app requesting the system service, are currently active in the foreground. By exploiting the hidden APIs, attackers can bypass all the operations in service helpers and silently invoke the system services without any foreground user interactions.

\begin{figure}[htbp]
\vspace{-10px}
\centering
\includegraphics[width=0.40\textwidth]{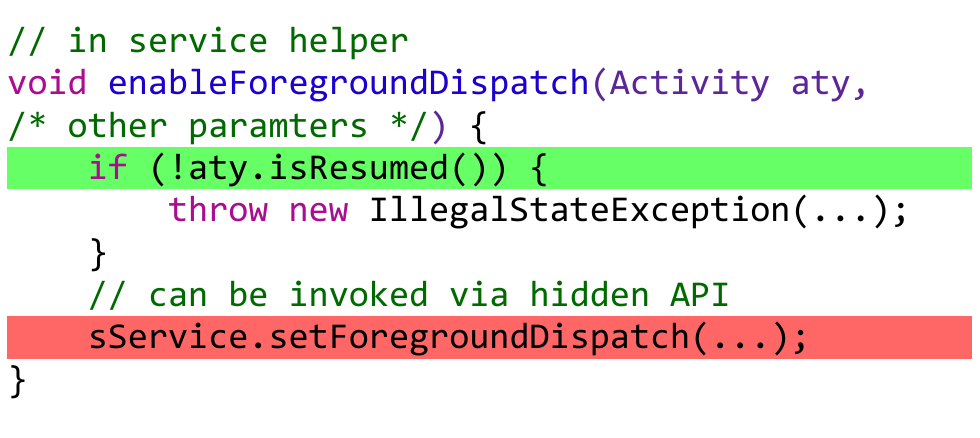}
\vspace{-15px}
\caption{Attackers can bypass the status checks in service helper.}
\label{fig:code2}
\vspace{-10px}
\end{figure}

\noindent \textbf{Evading the Validation of Caller's Status.}
To achieve the goal, service helpers verify callers’ status before the callers access the services. As shown in Fig~\ref{fig:code2}, NfcAdapter, which is the service helper associated with system service NfcService, confirms a caller’s status in \textit{NfcAdapter.enableForegroundDispatch()}. If the caller is not currently active in the foreground, the registration request for using NFC listeners will be rejected. Attackers can trigger this bug by directly invoking the \textit{setForegroundDispatch()} method via Java reflection and thus can invoke the NFC function without user's awareness.

\subsubsection{Crash the System Service and DoS Attack}
The attackers can bypass all sanity checks in service helpers and send illegal parameters to the service API. Then the system service helper may crash due to unhandled exceptions. Also, the attackers can send many IPC requests to the resource creating service APIs to exhaust the system resource and perform DoS attackers. Specifically, if the core services (e.g., \textit{ServiceManager}) are attacked and crashed, the whole system will shut down and reboot.

\begin{figure}[htbp]
\vspace{-10px}
\centering
\includegraphics[width=0.40\textwidth]{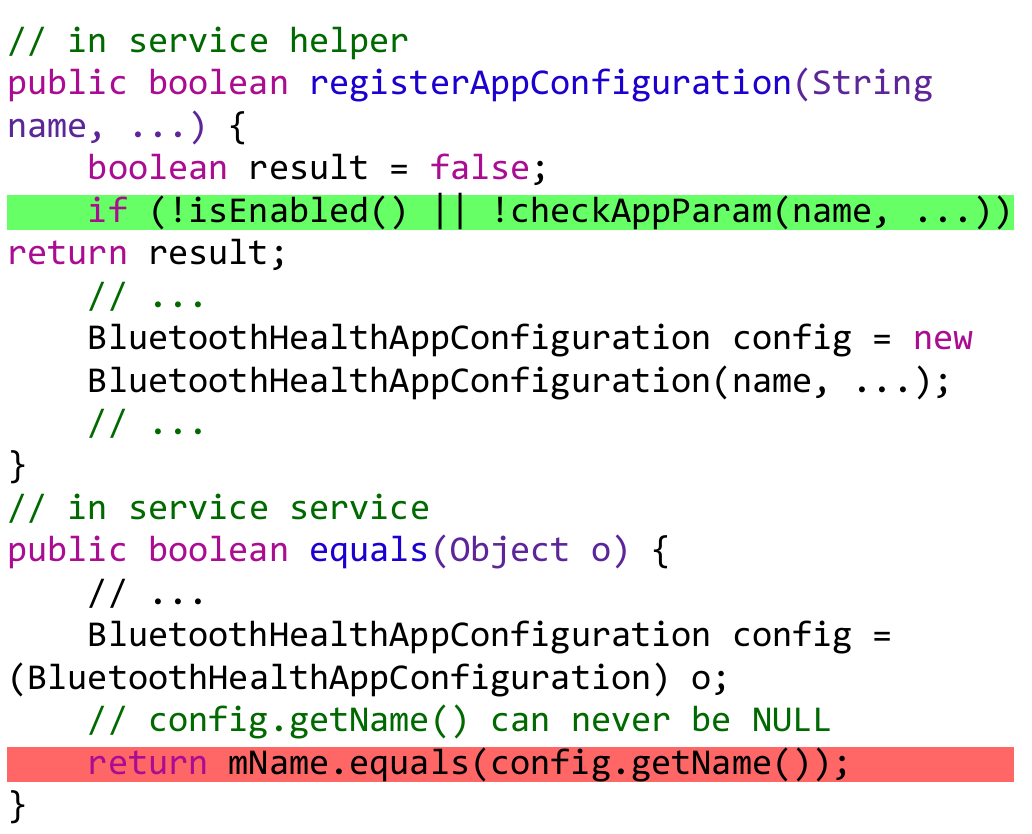}
\vspace{-5px}
\caption{The package name is not validate again in system service.}
\label{fig:code3}
\end{figure}

\noindent \textbf{Feeding Illegal Parameters.}
The \textit{HealthService} service, which provides health related Bluetooth service, contains a vulnerability that can be exploited by passing illegal parameters.
The method pairs, i.e., the service helper method \textit{BluetoothHealth.registerAppConfigeraton(String name, ...)} and the corresponding service method do not use the same method to validate the parameters.
The helper method checks the ``name'' parameter to make sure it is not null, whereas the service method does not.
The system service does not use the value of ``name'' parameter immediately in the IPC method.
Instead, it stores the value and uses it in \textit{BluetoothHealthAppConfiguration.equals()}. 
In \textit{equals()}, it assumes that \textit{config.getName()}, which returns the value of ``name'', can never be null (see Fig~\ref{fig:code3}). 
When a malicious app bypasses the service helper and registers a config with a null value in the ``name'' parameter, \textit{BluetoothHealthAppConfiguration.equals()} will generate a \textit{NullPointerExcetion}. 
Unfortunately, this system service method fails to handle the exception, and hence the \textit{HealthService} crashes. 

\begin{figure}[htbp]
\vspace{-10px}
\centering
\includegraphics[width=0.40\textwidth]{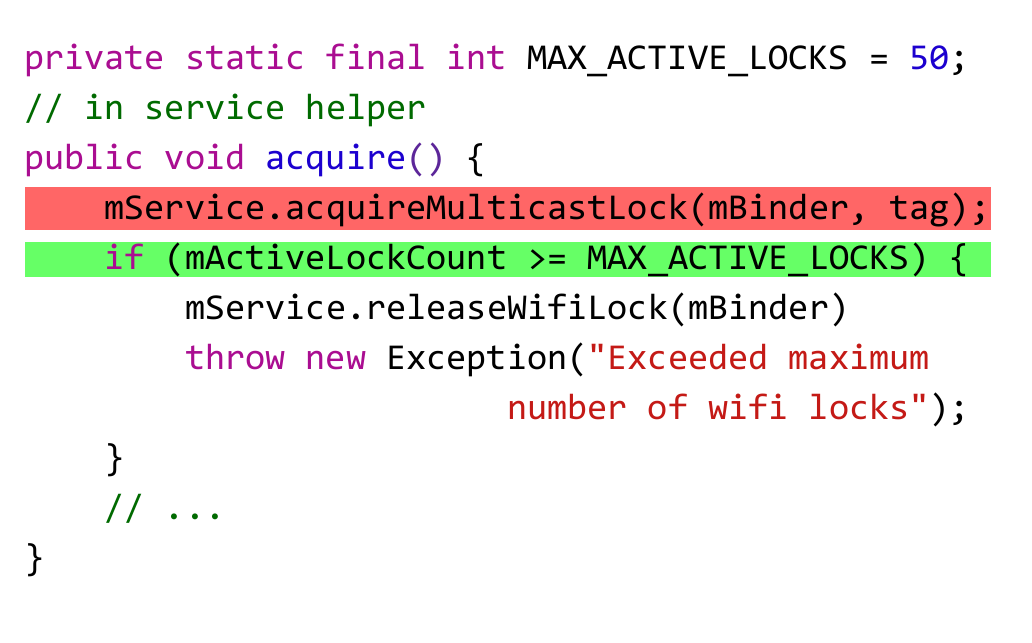}
\vspace{-15px}
\caption{The lock count is only checked in service helper classes.}
\label{fig:code4}
\vspace{-15px}
\end{figure}

\noindent \textbf{Sending Duplicated Requests to Exhaust System Resource.} As shown in Fig~\ref{fig:code4}, an app can get a notification of the clipboard changes by registering a listener to the system service \texttt{ClipboardService}.
\texttt{ClipboardManager}, which is the service helper associated with \texttt{ClipboardService}, only registers the service once to get such a listener after receiving the first request.  
After that, \texttt{ClipboardManager} maintains a local listener queue.
Any duplicated requests of registering listeners afterward will be only added to this queue locally.
When the clipboard changes, \texttt{ClipboardService} notifies its helper \texttt{ClipboardManager} about the change, and \texttt{ClipboardManager} notifies all the listeners in its local queue. 
Thereby, the system services only allocate resources for one request generated by the app, and can still notify the app when there is resource update. 

Normally, multiple IPC requests generated from an app can only access a system service during the app's lifetime and are only allowed to consume limited system resources of memories, CPU time, and file descriptors. Therefore, if a system service accepts many duplicated requests, the resources may be exhausted. To prevent resources from being exhausted by duplicated requests, service helpers constrain the frequency of local requests.  For resource requests, service helpers restrict the number of calls that an app can issue.  If the number of IPC calls to a target process exceeds a threshold, which can be treated as abnormal or unnecessary, the following calls will be dropped locally.  When duplicated requests are issued to register listeners, service helpers initiate an IPC call to the remote system service to make it only receive the first request of the app. Unfortunately, these security mechanisms enabled in service helpers can be easily bypassed if a malicious app directly invokes hidden APIs of the system services. As shown in Fig~\ref{fig:code4}, although the count is protected (see green line) by the service helper, the attackers can directly invoke the \textit{acquireMulticastLock} to create unlimited locks and make the system crash. The root causes of these attacks have been fully discussed in~\cite{ss_attack_1, jni_exhuast}.

\subsection{Google's Countermeasures in New Android Versions}
An obvious way to fix these bugs is to enforce the security check on the system services’ side. However, there are many corner cases, where the security enforcement cannot be easily accomplished in the services, such as the security check with specific status or parameters (see Fig~\ref{fig:code2}). Attacks exploiting these vulnerable services are constructed via hidden APIs and thus can be prevented by blocking these hidden APIs, which is adopted by Google. Unfortunately, a number of hidden APIs have already been used in many apps, and hence Google has to gradually include these APIs into the blacklist in several versions. Now we summarize the countermeasures adopted by Google in various versions.

\begin{figure}[htbp]
\vspace{-10px}
\centering
\includegraphics[width=0.40\textwidth]{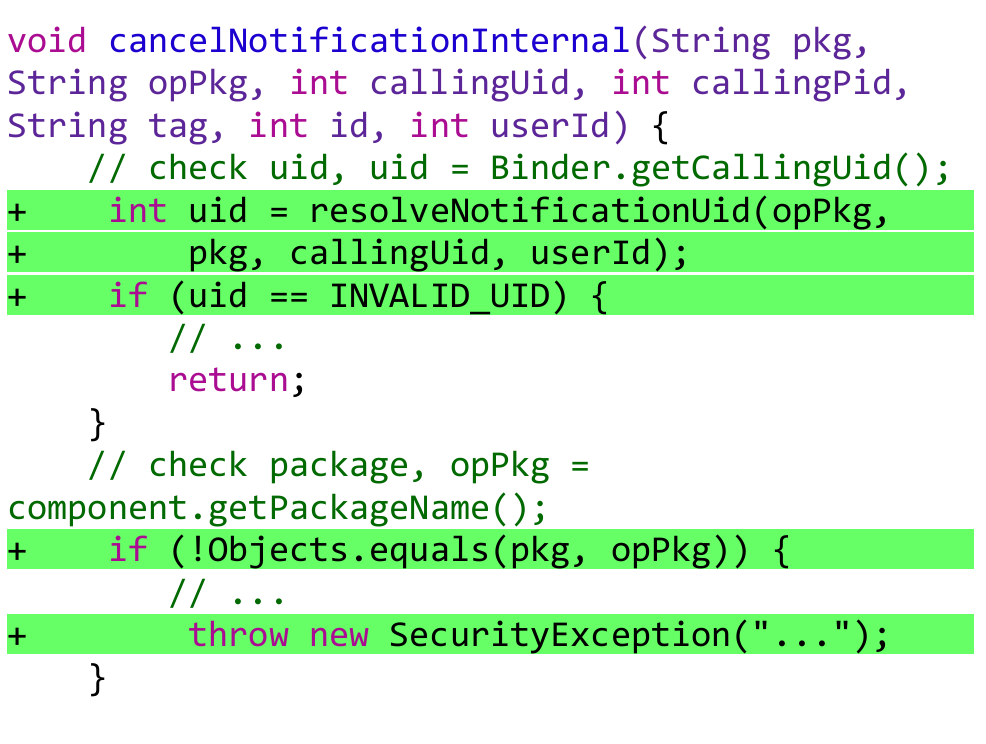}
\vspace{-15px}
\caption{Android 11's mitigate code for the Notification vulnerability.}
\label{fig:code5}
\vspace{-5px}
\end{figure}

\noindent \textbf{Ad-hoc bug fix.}
Google fixed the bugs one by one using ad-hoc methods for some specific problems. As shown in Fig~\ref{fig:code5}, in Android 11, Google fixed the cancel notification vulnerability by checking the real IPC  callers by obtaining its real uid from \textit{Binder.getCallingUid()} (in \textit{resolveNotificationUid} function). Another example is the IPC flood vulnerability. The root cause is that attackers create too many \texttt{Binder} objects in the core system process and make the JNI global reference table overflowed. Google  fixed the problem by modifying the native code and limited the number of the \texttt{Binder} objects since Android 8, but they failed to thoroughly fixed all the JNI global object leakage~\cite{jni_exhuast}.  

\noindent \textbf{Block The Hidden APIs.}
Google has set a blacklist filter in Java reflection methods to block the usages of these hidden APIs since Android 9~\cite{block_hidden_api}. However, there are still several ways to bypass the blacklist restrictions~\cite{bypass_blacklist}. If the developer downgrades their apps' by using older SDKs, their app can still use the \texttt{greylist} APIs in the new version because Google needs to keep compatible with apps using these APIs. To address this problem, Google Play does not allow apps with such APIs to be published now.

\noindent \textbf{Restrict Every System Services API with Common Enforcements.}
The ideal solution to mitigate this problem is to automatically add a common security mechanism in every IPC entrance method of the system services during the AIDL interface generating. The system service can obtain the identity of Apps such as \texttt{uid} and package name from the \texttt{Binder} and double-check the parameters from service IPC methods. In this way, system services do not need to trust the identity from the client app side and can get real identity. For each IPC method, the system service needs to check the permission of the calling \texttt{uid} via \texttt{checkCallingPermission()}. Currently, Google does not apply these common caller checks to all the IPC functions but manually added checks in many places.

\subsection{Evade the Hidden API Blacklist}
\label{sec:evade}
Though Google has taken measures to block the Java reflection or directly calling to the Hidden APIs. Developers or hackers have evaded the hidden API blacklist as quickly as each new Android version~\cite{bypass_blacklist, bypass_blacklist_12} is released. 
Since the framework code is run in the same process of user code, the addresses of all the Java and native methods of framework can be obtained by the attackers. Thus, the attackers can obtain the reference of the blacklist policy setting functions (\textit{VMRuntime.setHiddenApiExemptions}) from the app's memory space and just call it to remove the blacklist~\cite{bypass_blacklist_12}.

The non-SDK APIs as a part of the Android framework are widely used by other code of the \textit{framework.jar}. Thus, it is inadvisable for Google to prevent the app code from calling the hidden APIs by checking app identity in the system services' side, e.g., disallow user app to invoke the hidden API IPC methods. As a result, Google has to implement  blacklist restrictions in the framework code and run in the same process with the user app which leads to a dilemma, the app process disallows some part of its code to access the hidden APIs but other parts can access these APIs. Unfortunately, Android does not support in-app privilege isolation~\cite{app_isolate} and the remote services cannot distinguish whether the IPC requests are launched by the framework code or app user code. Thus, the attackers can eventually break the blacklist from the memory via native code. 
\section{Identify Vulnerable Service API}
\label{sec:hazards}
To automatically identify the vulnerable service APIs in different Android versions, we propose a new static analysis tool \ourwork. In this section, we introduce the design of \ourwork. For each module, we illustrate its detailed design and its difference with existing approaches. In particular, we highlight the methodology and technique advancements of \ourwork comparing to our previous approach~\cite{Gu16}.

\begin{figure*}[t]
\centering
\includegraphics[width=0.8\textwidth]{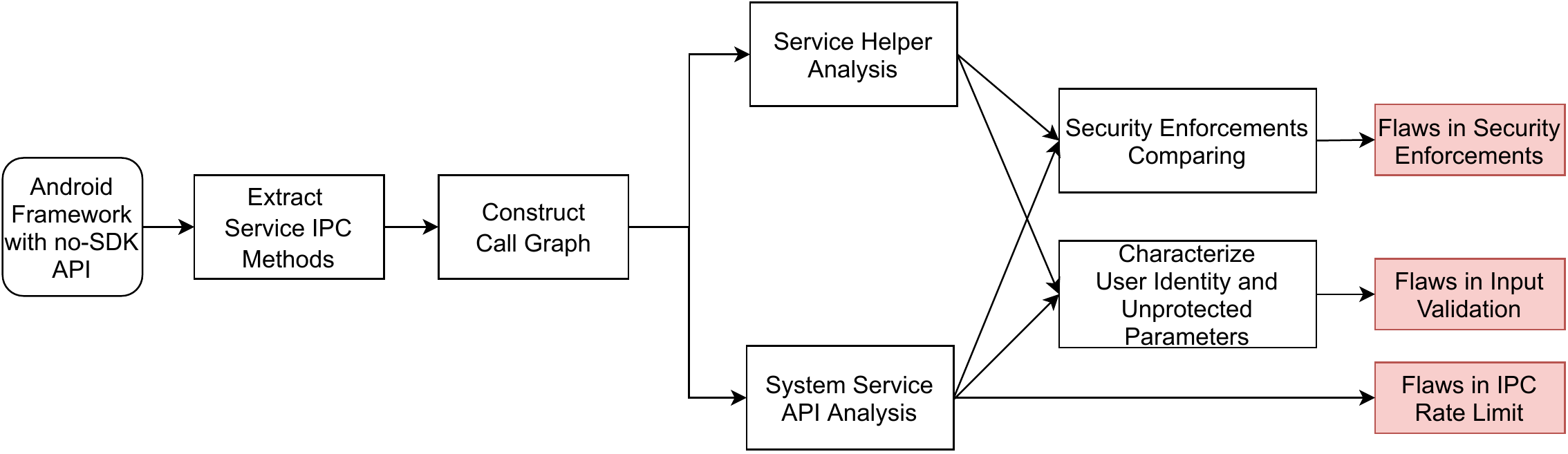}
\vspace{-10px}
\caption{Overview of our approach.}
\label{fig:method_architecture}
\vspace{-15px}
\end{figure*}

\subsection{Approach Overview}
\ourwork statically analyzes the Java bytecodes of the Android framework and finds out the inconsistent security mechanisms between service helpers and system services. It operates in four steps, as shown in Fig~\ref{fig:method_architecture}. First, we extract the full \textit{framework.jar} which contains all the no-SDK hidden service APIs as the analysis target. Second, we extract service IPC methods and classes, including both service helpers and system services.   For each IPC method, we find out which service helper classes invoke it and then associate these classes with the corresponding system service class, which implements the IPC methods. Third, we build a method level call graph for each IPC method, which begins with a service helper method and ends with the target IPC method. Then we mine security enforcement rules from the calling paths from the call graph which indicates how a service helper method warps an IPC method. Thus, we can gain all the security enforcements on the service helper’s side. For the system service method, we collect the security mechanisms by analyzing the methods called around the IPC.stub method. Finally, we enumerate all IPC methods and compare whether the security mechanisms enforced by service helpers are consistent with that in the corresponding system service. If they are not consistent, we treat it as a potential vulnerability that may be exploited by attackers through bypassing the service helper. Specifically, we have implemented several analysis patterns to identify different types of security enforcements, e.g., patterns for identifying sanity checks, caller identity, IPC flood. After finding out the potential vulnerability, we need to manually evaluate and confirm it since not all potential vulnerabilities can be exploited in practice. 

It is non-trivial to identify the inconsistent security enforcements in hidden IPC methods of system service. There are three significant challenges to achieve this. First, it is not easy to find out all corresponding service helper classes associated with the identified system services.  In particular, the internal and hidden APIs~\cite{ss_attack_2} that can be invoked by third-party apps through Java reflection is not included in the official Android SDK. If we intend to identify all service helper classes that third-party apps can access, we need to obtain a full version of \textit{framework.jar} contains all these APIs while in different Android versions the compiling of framework changes dramatically. Second, since system services and the corresponding service helpers use different mechanisms to verify the parameters included in the requests, it is challenging to identify such enforced security mechanisms accurately.  Finally, it is not easy to identify and compare security mechanisms in system service methods and the corresponding service helper methods. The security mechanisms in service helpers and the corresponding system services are normally implemented with different methods. Service helpers run in the same processes with the third-party caller app, whereas system services run in a separate system process. Therefore, we cannot simply compare the source code of security mechanisms in system services and their corresponding service helpers.

To address these issues, we need to improve our approaches and accommodate the changes of new Android versions. Compared to our previous approach~\cite{Gu16}, we have rewrote the whole tool for higher precision, better performance, and more features for detecting new vulnerabilities. Also, we clarify the discrepancy between our works and existing approaches~\cite{acedroid, invetter}.

\subsection{Extracting Services IPC Methods}
In this step, we mine out the service IPC methods and pair the service helper methods (which invokes the \textit{IPC.proxy} methods) with the corresponding system service methods (\textit{IPC.stub} method). Thus, we can further analyze the security enforcements of both sides of a IPC method.

Here we use a similar approach to our previous work~\cite{Gu16} to extract the service IPC method pairs. For system service classes, we can identify them by analyzing the services register invokes in \texttt{com.android} package. Since all services are registered in the service manager via \texttt{addService} or \texttt{publishBinderService}, it is possible to obtain these system services effectively via analyzing if the services have registered themselves using aforementioned methods. Then we need to obtain the IPC methods of these system services, which can be extracted from the \textit{IPC.stub} classes. Note that some native services cannot be analyzed and need to be manually verified in the vulnerability analysis phase.

After this, we extract the service helper methods from the \texttt{android} package. We consider a class as a service helper only when it has one or more methods that invoke an IPC method to invoke the system service. By enumerating all public classes in the Android SDK, we can filter out the classes containing service IPC method calls which can be verified by the callee’s java class type, as a service IPC class should have implemented both \texttt{IBinder} interface and \texttt{IInterface}. Thus, we can identify which service they invoked by analyzing the class hierarchy via CHA~\cite{cha_paper} analysis to obtain the IPC class’s real type. Based on this rule, service helper classes can be identified by enumerating all the fields and method invokes and verify whether the fields or methods are from service IPC classes. If an identified class is a nested class, an inner class, a local class, or an anonymous class, the top-level enclosing class will be treated as the service helper class.

Finally, we associate the service helper method with the system service method based on the common service IPC method they invoked, and then save the two methods as a pair for security mechanism analysis. To achieve this, we need to enumerate all the service helper methods to find out the IPC client method (\textit{IPC.proxy}) they used and then enumerate all the service methods to find out which implement the IPC server side (\textit{IPC.stub}). In this way, we can extract a system service API to one or more system helper methods that invoke it.

\subsection{Constructing Call Graph for IPC Methods}
\label{sec:build_callgraph}

We use context-insensitive control flow analysis to identify security enforcements and detect vulnerabilities for each service IPC method. The most important is to build an effectiveness call graph with correct object types.

\textbf{\textit{Improvement 1: Building accurate call graphs.}} Rather than directly using the Soot’s full call graph, we build a method level call graph that only focuses on the edges between service helper method and the IPC API to improve the performance. Moreover, we employed multiple technologies (e.g., EdgeMiner~\cite{edgeminer}) to improve the virtual calls, implicit calls used in Android framework to correctly resolve the callback methods or inherit methods. Thus, more security enforcements can be discovered on the calling paths.

Our method level call graph is built via breadth-first searching on the fly. Starting from each service helper method, we add each invoked statement into the worklist. Finally, if finding a service IPC method is invoked, we save the whole call graph chain between this method pair. In this way, we can build a simplified call graph for each service IPC method which only includes the edges that are useful to our analysis. Then, we perform the same approach to build the call graph for system services’ IPC methods. Our approach is much more efficient than the Soot's default approach which needs to perform whole program analysis as we only add limit call graph edges around the IPC methods. 

To improve the accuracy of the call graph, we take various approaches to precisely resolve the virtual invokes and implicit calls to get the methods that are actually called on the path. For normal dynamic dispatch, we utilize CHA ~\cite{cha_paper} call graph building algorithm to get a possible method set that may be called by the virtual invoke's callee. Fortunately, the hierarchy of Android service classes is not very complicated, especially for service IPC method, with only one or two inherit levels. However, Android SDK contains many implicit calls, e.g., Binder IPC calls, Messenger, Handler, and other callbacks such as UI OnClick events. To address this issue, we take advantage of the callback results of EdgeMiner~\cite{edgeminer}. If one of the interfaces of a class is in the callback set, we need to retrieve the implemented method of the callback from the subclass of the interface.

For each system service method associated with a service helper method, we only need to analyze all the invoke statements in the entry method as they can be exploited via Binder IPC directly. Finally, for each IPC method, we need to check the security mechanisms in the service helper method's call graph and invoked methods of the system service method.

\subsection{Identifying Security Enforcements}
\label{sec:indentify_enforcement}
To discover the inconsistent security enforcements between the service helper method and system service method, we first need to examine the presence of security mechanisms in both system services and service helpers by extracting and comparing the patterns of different security mechanisms. The presence of security mechanisms in system services and their helper classes need to be analyzed separately with different approaches since they may have various implementations for the same security mechanisms. 

\subsubsection{Mining New Security Mechanisms}
\label{sec:npl_approach}
In our previous approach~\cite{Gu16}, we have labeled some of the system security mechanisms including methods for getting user identify like \texttt{getPackageName} shown in Table~\ref{tab:id_method_for_sh}, Table~\ref{tab:check_method_for_ss}, and methods for checking user permission, e.g., \texttt{enforceCallingOrSelfPermission}. However, the method set is incomplete and failed to include all the access control security checks as different services may use various implementations to perform security and argument checking. Hence we use the natural language processing (NLP) approach to automatically discover such security checking mechanisms with aliasing method names different from the common security enforcements~\ref{tab:check_method_for_ss}. Then we apply this approach to both service helpers and system services to new mining new security mechanisms other than the standard system security mechanisms before the vulnerability detecting.

\textbf{\textit{Improvement 2: Discovering more security enforcements.}} We propose NLP approach to mine the customize permission checking functions with alias names. Thus, we can obtain more security enforcements on the system services' side and reduce the false positives.

To identify these customized security mechanisms methods using synonyms names in service code, we leverage associate rule mining~\cite{associate_rules} analysis and word similar analysis to mine out new security mechanisms keywords. A similar work Invetter~\cite{invetter} focuses on insecure inputs that are not properly validated by the system service. Our main difference to Inverter is that we concentrate on the permission checking function's method names. At the same time, Invetter focuses on the param name to find out sensitive input parameters. And Invetter's approach does not work in our situation, as our analysis target is framework classes compiled from the Android source code. In our version of framework.jar, origin names of local variables and function parameter names are lost due to compile optimizing. We try to retrieve the local fields and parameters' raw names but only get temporary names like \texttt{str1} and \texttt{paramInt}, only member names and method names of a class stay the same. On the contrary, the function name of the security enforcements keeps unchanged after compiling.

Our approach does not rely on Java code's exact parameter name so that our approach can work on any compiled code, even with variable obfuscation. Notice that service helper methods perform extra security checking before the IPC call, and system service may perform security checking again at the IPC interface implementation methods entrance. Around these IPC methods, we can collect a set of methods called before or after each of the IPC methods. Some of them check the user's parameters and privileges, which might be the actual implementation of security enforcements. By analyzing the method level callgraph, we can extract all these methods with Soot Stmt~\cite{soot} analysis. For each method of the call chain, we retrieve all the Stmt of its method body and check if the Stmt is invoked Stmt or definition Stmt and then we resolve the real type and name of these variables functions by point-to analysis. After extracting a method name list for each service IPC method, we get a method name list collection containing the method calls in every IPC call chain. Specifically, we manually obtain keywords from the user identity accessing API in Table~\ref{tab:id_method_for_sh} and the user identity-checking API in Table~\ref{tab:check_method_for_ss}. And we use the keywords to filter out these method name lists, which do not contain the keywords and use the left items as input for the associate rule mining. Thus, from the mining results, we can find other high-frequency methods that use these user identity methods, and most of these high-frequency methods directly call the user identity APIs or provide arguments for user identity methods. The mining results are shown in ~\ref{tab:associate_methods}.

We can gain most high-frequency method names from the association rules mining result, which are always used together with user identity APIs, and we regard them as new security mechanisms methods. All these method names and the user identity APIs are used to identify security enforcements in the following analysis.

\subsubsection{Identifying Enforcements by Backward Dataflow Analysis}
\label{sec:dataflow}

The accuracy of identifying enforcements in both server helpers and system service is significant to tool precision since the miss cases in server helpers can lead to false negatives and the miss cases in system service can increment the false positives. We perform precisely analysis to check if the parameters are actually checked by these enforcement methods.

\textbf{\textit{Improvement 3: Precisely tracking the IPC parameters with dataflow analysis.}} By employing the dataflow analysis for the IPC parameters, we can verify if the fake identity (see Fig~\ref{fig:code-1-2}) and illegal parameters (see ~\ref{fig:code2}) are actually validated by the system service by further tracing the dataflow of these parameters while our previous approach~\cite{Gu16} can only checks if there are enforcements methods in the entrance level of a system service API. Thus, our new approach can detect if these parameters are validated in other sanity check functions (e.g., \textit{checkParams}) in a deeper level of the calling path and mine more security enforcements in system services.

We perform backward dataflow analysis to track the parameters of service IPC methods and collect the methods that use these parameters before IPC call in the call chain. For the methods using these parameters, we also perform the Stmt analysis in these methods to determine the usage of security mechanisms related methods. The detailed approach of Stmt analysis is to enumerate each Stmt in the Soot method body and find out the invoked  Stmt and then extract the method name and parameter information. For a reference type, we obtain their real name and the Java type via the points-to analysis. We track the initialization and assignment of each Java reference value in function level and regard the same references to a value as the same pointer.  In this way, we verify if the local fields are used by the user identity accessing methods and further analyze if these local fields are passed to IPC methods or security enforcement methods. This approach is also applied in the system service methods to check if there is any security enforcement. 
We use dataflow analysis and points-to analysis to identify whether the parameters of IPC methods are protected by sanity check functions that validate the parameters and throw exceptions (see Section~\ref{sec:find_input_check}) in service helpers.  
We trace the dataflow starting from the IPC methods and analyze the backward dataflow of the corresponding parameters by using  \texttt{BackwardFlowAnalysis} in Soot. Thus, we can identify all security checks for the parameters which are conveyed to the service IPC methods.

\subsubsection{Identifying Security Mechanisms in Service Helpers}
\label{sec:identify_security_checks}
We use different patterns to identify the presence of various security mechanisms in service helpers.

\noindent{\bf Identifying Parameter Validation.}
\label{sec:find_input_check}
We can capture this type of security enforcement mechanism by analyzing if any exceptions will be triggered by handling illegal parameters. The key observation is that the Binder framework of the Android system does not handle all exceptions though most of them incurred by illegal parameters will be captured and handled. For example, 
the six most common exceptions are well handled in the system services defined by AIDL, including \texttt{BadParcelableException}, \texttt{IllegalArgumentException}, \texttt{IllegalStateExce-}\newline\texttt{ption}, \texttt{NullPointerException}, \texttt{SecurityExcept-}\newline\texttt{ion}, and \texttt{NetworkOnMainThreadException}~\cite{parcel_exception}.
If there are any other exceptions, the framework will re-throw them as \texttt{RuntimeExceptions}.
We observe that exceptions that are thrown by system service IPC methods and caught by the Binder framework are passed to the IPC callers through \texttt{Parcel.writeException()}. Thereby, the caller could handle these exceptions in its own process, which ensures that the service will not crash. We find that if a parameter of IPC methods in the service is used out of the methods (i.e., it is used in asynchronous handler or stored for later access), it may lead to failures of handling the generated exceptions.  
Hence, we can identify the security mechanism of parameter validation by analyzing if illegal parameters are used out of IPC methods. 

To achieve this, We adopt \texttt{def-use} analysis~\cite{aho1986compilers} and backward dataflow analysis to identify all parameter validation mechanisms. The \texttt{def-use} analysis links each variable definition with that is referred so that we can identify if a variable, i.e., an input parameter, is referred to a validation process. We can identify validation mechanisms if the following two conditions are met:  
(i) We check whether input parameters of a method are used in boolean expressions or whether they are used in other methods that return boolean values; (ii) If the parameters are indeed associated with boolean values, we further verify whether the boolean values are used in conditional statements that contain statements with early returns or thrown exceptions. If these two conditions are met, the method includes input parameter validation. To reduce the false positives of identifying the validations, we use the dataflow and point-to analysis (see Section~\ref{sec:dataflow}) to precisely check if the parameters are actually checked among different procedures. If parameters stored for later access can be accessed by the classes in the IPC methods or the fields of the outer classes, e.g., a parameter is accessed and included in a collection of the field, it indicates that exceptions triggered by input parameter validation indeed exist. Moreover, we identify all possible asynchronous handlers such that we can accurately find if a parameter used in an asynchronous handler will be validated. Thereby, we can identify the parameter validation mechanism in service helpers.

\noindent {\bf Identifying Caller Status Validation.}
Caller status validation is similar to the input parameter validation.  We identify the mechanism of caller status validation by (i) analyzing if there exists APIs that return callers' status and (ii) verifying if these APIs are used in conditional statements. If these two conditions are met, a caller status validation mechanism in the service helper is identified. 

\noindent{\bf Identifying the Process of Passing Caller's Identity.}
\label{sec:Identify the process of passing the caller's identity.}
We identify this mechanism by searching if any method processing identity information is executed before the IPC method of a target service. Normally, apps can provide different types of identities. As we observed, there are seven types of identity information in Android, i.e., package name, \texttt{uid} (i.e., Linux user identifier), \texttt{pid} (i.e., Linux process identifier), \texttt{gid} (i.e., Linux group identifier), \texttt{tid} (i.e., Linux thread identifier), \texttt{ppid} (i.e., Linux parent process identifier), and \texttt{UserHandle} (i.e., representing a user in Android that supports multiple users).
Each of them can be obtained by calling different methods. These user identity accessing methods are summarized in Table~\ref{tab:id_method_for_sh}. Methods from the associate rules mining results such as \texttt{getCallingUid} and \texttt{getCurrentUser} are also added to the user identity accessing method set. Suppose any method included in a service helper tries to access user identity before an IPC method of the call graph's target service. In that case, the service helper may validate the caller's identity and then pass the verified identity to the corresponding system service.

\setlength\extrarowheight{3pt}
\begin{table}[ht]
\caption{Methods used by service helpers to obtain caller's identity.}
\centering
\begin{tabular}{ | c | l | }
\hline\hline
{\textbf{Identity Type}} & \multicolumn{1}{|c|}{{\textbf{Method}}}  \\ [0.5ex] 
\hline\hline
\multirow{3}{*}[-2pt]{Package Name} & Context.getPackageName()\\ \cline{2-2}
             & Context.getBasePackageName()\\ \cline{2-2}
             & Context.getOpPackageName()\\ \hline
\multirow{3}{*}[-2pt]{UID}          & Process.myUid()\\ \cline{2-2}
			 & Process.getUidForPid()\\ \cline{2-2}
			 & Context.getUserId()\\ \hline
\multirow{3}{*}[-2pt]{PID}           & Process.myPid()\\ \cline{2-2}
			 & Process.getPids()\\ \cline{2-2}
			 & Process.getPidsForCommands()\\ \hline
\multirow{2}{*}[-2pt]{GID}           & Process.getGidForName()\\ \cline{2-2}
			 & Process.getProcessGroup()\\ \hline

\multirow{2}{*}[-2pt]{PPID}          & Process.myPpid()\\ \cline{2-2}
			 & Process.getParentPid()\\ \hline
TID           & Process.myTid()\\ \hline
UserHandle   & Process.myUserHandle()\\ \hline\hline
\end{tabular}
\label{tab:id_method_for_sh}
\end{table}

\noindent {\bf Identifying the System Environments Checks.} 
We capture this security mechanism by identifying if a service helper invokes the corresponding system service according to the boolean result returned by an IPC interface of the system service. We enumerate all IPC interfaces in the system service returning boolean values since the interfaces that include the methods of evaluating system environments normally return boolean values. In general, the interfaces that evaluate system environments verify if a hardware device, e.g., a fingerprint reader, exists before using the device, or verify if a function, e.g., accessing the wallpaper, is disabled before invoke the function, and then return boolean values. We further check if the boolean statement includes a method of the corresponding service helper. If it is true, it indicates that the service helper may invoke a method in the system service accessing the system environment according to the boolean results of evaluating the environment.

\noindent{\bf Identifying the Constraint of Duplicated Requests.}
\label{sec:Constraining Duplicated requests.}
Service helpers adopt two mechanisms to constrain the number of duplicated requests delivered to the system services according to the types of requested resources (see Fig~\ref{fig:code4}) so that they prevent different resource consumption incurred by duplicated requests from apps. The first mechanism is to constrain the number of requests that an app can issue for resource access. If the total number of requests exceeds a threshold, the subsequent requests will be ignored. To identify the existence of this mechanism, we search all methods in service helpers and locate the condition statements with integral constant expressions.  If such a condition statement is located after the entry of the corresponding service helper but before the IPC calls to the service in the call graph, there is a high probability that the statement is used to check duplicated requests, which is similar to input parameter validation. The second mechanism is to constrain the number of duplicated requests to register listeners. Usually, a service helper method accepts a listener as its parameter, and saves the listener to a local list. For example, the service helper method, \texttt{EthernetManager.addListener(Listener listener)} saves the parameter listener to \texttt{ArrayList<Listener> mListeners}. If it is the first registration request, the helper will register in the remote service via IPC.Otherwise, the service helper method only adds the request's listener to the local list. When the service helper receives an update from the service, it dispatches the update to all listeners maintained in that list. We can capture this type of mechanism by identifying the code maintaining the listener lists.

\subsubsection{Identifying Security Mechanisms in System Services}
The approaches to identifying security mechanisms in service helpers (see Section~\ref{sec:identify_security_checks}) can also be applied to identify the security mechanisms enforced in system services. However, we cannot directly adopt it to identify security mechanisms in system services because of the following difference between systems services and service helpers. Firstly, service helpers run in the same process with the caller while system services do not. 
Service helpers can directly obtain the caller's identity via methods in Table~\ref{tab:id_method_for_sh}.
However, system services need different APIs to obtain the information about the caller and check the caller's properties, since they run in system processes that are separated from the caller processes. 
For instance, system services use \texttt{Binder.getCallingUid()} and \texttt{Binder.getCallingPid()} to obtain the callers' identities instead of the methods listed in Table~\ref{tab:id_method_for_sh}. 
Secondly, system services need to validate app identities and verify whether the calling app has privileges to perform sensitive operations, which is not required in the service helpers.
Fortunately, we find that system services heavily rely on the functions provided by \texttt{AppOpsService} to perform validation. 
For example, \texttt{AppOpsService.checkPackage(int uid, String packageName)} checks whether the input package name actually belongs to the given uid, and \texttt{AppOpsService.checkOperation(int code, int \texttt{uid}, String packageName)} checks whether the \texttt{uid} has the privilege to perform the sensitive operation indicated by the \texttt{code}. Therefore, we can identify all functions provided by \texttt{AppOpsService} and use these functions to identify the security mechanisms in system services. These system' default permission checks functions in system services are listed in Table~\ref{tab:check_method_for_ss}. We mine extra user identity checking methods do not use the default functions via the NLP approach(see Section~\ref{sec:npl_approach}), such as \texttt{enforceUriPermission}, \texttt{checkReadPermission} and \texttt{enforcePermission} (shown in Table~\ref{tab:associate_methods}).

Since the adversary intends to exploit the services that do not have the same enforcements with their associated service helpers, it is crucial to distinguish whether different forms of enforcements are equivalent. To improve the precision, we further verify the enforcements that are actually applied to the same parameters of the IPC methods, not for other purposes. By focusing on the security enforcements around the IPC parameters and analyzing the reference of these parameters to discover how these parameters are eventually validated (by sanity checks or permission checks), we have significantly reduced the false positives.

\subsection{Detecting Possible Vulnerabilities.}
The final step is to capture the service helper bypass vulnerabilities by examining whether the security mechanisms in the method pairs (i.e., the service method and the corresponding service helper method) we identified in Section~\ref{sec: Find Service Helper Classes} are consistent. We identify the inconsistency of security mechanisms in these methods by leveraging the prebuild call graph and detecting their code features in the graph. 
\subsubsection{Identifying Enforcement Inconsistency}
Our tool can automatically perform various types of verification. For the inconsistent parameter checking, we can simply compare if a service helper verifies more parameters than that in the corresponding system service. We extract the parameters validated in each party of the method pair into two sets. The parameters validated in the service method are denoted as set $S$, and those in the service helper method are denoted as set $H$. If $S$ is \textit{not} the superset of $H$, which means the helper checks more parameters than the service, we use the dataflow analysis to check if there are parameters that are checked by service helper classes but not checked by the system service. In this way, we can find out the parameters of service IPC methods that are not well protected and regard them as vulnerable APIs. For the vulnerabilities of inconsistent security enforcement, we first analyze how service helper classes use the service IPC methods. If an IPC method in service helper classes is protected by security checks, we can confirm which security enforcement this method should implement. Then we check if the system service side also implemented the corresponding security enforcement to verify the existence of the protection mechanism. 

Moreover, we utilize permission checks to reduce false positives of identifying the bypass vulnerabilities. We filter out system service interfaces protected by high-security level permissions, i.e., the \texttt{signature} and \texttt{signatureOrSystem} levels, since these interfaces cannot be accessed by unauthorized apps and thereby cannot be exploited. We use PScout~\cite{pscout} to obtain interfaces with security level permissions and use the information to detect the vulnerable system service interfaces. Thereby, we can accurately identify the service helper bypass vulnerabilities incurred by such vulnerable system services.

Finally, potential vulnerabilities can be captured by comparing whether the security mechanisms in each method pair are consistent. If they do not match, it means that the system service may be possibly exploited by bypassing the corresponding service helpers (more specifically, bypassing the inside security mechanisms in the helpers). Since not all vulnerabilities are exploitable, we manually confirm the vulnerabilities by constructing real attacks.

\subsubsection{Identifying Exploitable Vulnerability}
As Google uses \texttt{blacklist} and \texttt{greylist} to constrain these hidden APIs, we need to check if third-party apps can actually invoke the vulnerable API. For versions after Android 9, we scan the \texttt{blacklist} to statistics how many vulnerabilities APIs are covered by these lists. The APIs in \texttt{greylist} can still be invoked by downgrading the app's SDK version in the build configuration. However, with the \texttt{greylist} and \texttt{blacklist} changes in every version, we still need to run the app on the specific device to verify if the vulnerabilities can be triggered. To estimate how many vulnerable APIs can be exploited in the worst case, we need to consider if these vulnerabilities can be exploited when the attacker evades the \texttt{blacklist} restriction and can abuse all the hidden APIs.






\section{\ourwork Implementation}
\label{sec:implementation}
We implement the static analysis tool based on Soot~\cite{soot} framework and the source code is open-sourced online\footnote{https://github.com/fripSide/ServiceAudit}. In this section, we elaborate on some technical details of the implementation.

\noindent{\bf Extract Android SDK Jar.} To analyze the no-SDK APIs, we need to obtain a full \textit{framework.jar} that contains all the code of the Android framework including all the service helper classes and system service implementation classes in it. A high fidelity way to obtain the classes without dead code is to extract the framework from a live Android Image. In Android 6 to 9, we need to extract the full SDK jar from various Android images which use \texttt{.odex} and \texttt{vdex} for the new Android Runtime (ART). Then we need to convert the \texttt{.odex} and \texttt{.vdex} to \texttt{.jar} files via oat2ex~\cite{oat2dex} and dex2jar~\cite{dex2jar} tools. However, the oat2ex~\cite{oat2dex} does not work for Android 10 to 12. To solve the problem, we build the AOSP from source and extract the corresponding classes from the compiling intermediate files distributed in different directories. To ensure all the methods are included, we need to manually add all the corresponding packages of system service and service helpers in \textit{"framework/base"} directory of the AOSP source code.

\noindent{\bf Mine extra security enforcements}
We use FpGrowth~\cite{fpgrowth} algorithm to perform associate rules mining. Our approach has been illustrated in Section~\ref{sec:npl_approach}, and we implement it with the following steps: First, we extract method names from the call chain of the services IPC methods' call graph and use it as the initial dataset for FpGrowth. Second, we manually set a sensitive API list which contain the system's sensitive APIs shown in Table~\ref{tab:id_method_for_sh} and Table~\ref{tab:check_method_for_ss}. When we build the initial FpTree~\cite{fpgrowth}, we check and set the items in the sensitive API list with a frequency of 1000 for two purposes: to make sure the sensitive APIs presence in each result set and to identify the items in results are associate with which sensitive API. And the minimum support frequency is set to 3 to filter out the less high-frequency items. Finally, we collect several lists of high-frequency items from the output of FpGrowth and use some keywords such as enforce, user, and check to perform word synonyms analysis to filter out these methods with names which unlikely to be security checking methods. The items remained are regarded as extra security enforcement functions.
{
\setlength\extrarowheight{3pt}
\begin{table}[ht]
\tinyv
\caption{New Security Mechanisms Methods Mining Results.}
\centering
\begin{tabular}{|c|l|}
\hline \hline
\textbf{Keywords}      & userid, uid, pid, identity, package,  \\
              &	enforce, permission\\ 
\hline\hline
\textbf{Identity Access}    & getCallingUid, getUuids, getCurrentUserId,  \\
\textbf{Methods} 		   & getPackageInfoNoCheck, getCallingPid, \\
& getDeviceOwnerPackageName,getUserId, \\
& getEffectiveUserId, getCurrentUserId, \\
& getCallingUserId, getPermissionPolicy\\
\hline
\textbf{Identity Enforce}  & checkOp, checkOperation, \\ 
\textbf{Methods} 		   & checkCallerIsSystemOrSameApp, \\
& enforceAccessPermission, checkPermission,\\
& enforceChangePermission,  \\
& enforceConnectivityInternalPermission, \\
& enforceCallingOrSelfPermission, \\
& enforceCallingPermission, \\ 
\hline \hline
\end{tabular}
\label{tab:associate_methods}
\end{table}
}


\noindent{\bf Verify and Exploiting the Potential vulnerabilities.}
After obtain the potential vulnerabilities, we need to verify these potential vulnerabilities. Traditional fuzzing tools such as AFL~\cite{afl} cannot use in our case as service IPC methods need to invoke via Java code or C++ code, and the method parameters need to be filled correctly. Customize fuzz code need to be generated for the fuzzing tests. We tried to parse the definition of IPC methods and generate invoking code via Javapoet~\cite{javapoet} for automatically test. And we obtain the information for the crash, soft reboot, or permission usages from Logcat. Unfortunately, due to the complicated IPC methods parameters, some of the generated code is failed to compile. The verification works need manual works to read the source code and write test apps to verify in real devices.
\section{Vulnerability Results}
\label{sec:Findings}
We prototype our approach which is presented in Section~\ref{sec:implement} and apply it to analyze the AOSP 6.0.1. We find a large number of service helper bypass vulnerabilities that incur serious security issues. In this section,  
we analysis the accuracy and effectiveness of our static analysis tool and then summarizes our findings and evaluate whether Google's countermeasures work in various versions. In particular, we compare the detection accuracy of our approach with our previous approach (\texttt{Result$_0$})~\cite{Gu16} as well as other existing works. Moreover, we construct real-world attacks to exploit a representative vulnerability existing in each security mechanism.

\subsection{Summary of Results}
We experiment on a laptop computer with a 6-core Intel i7-8750H CPU and 16GB memory, running 64-bit Ubuntu 18.04. The whole analysis for AOSP 6.0.1 extended SDK can be done in less than 15 minutes. We use \texttt{ServiceAudit} to check 4 Android versions range from Android 6 to Android 10. Unfortunately, we failed to extract the full SDK Jar from Android 10 and Android 11. The analysis results are shown in Table~\ref{tab:my-res}. \texttt{ServiceAudit} can find out hundreds of potential vulnerable APIs among various Android images.  To measure the accuracy and efficiency of \texttt{ServiceAudit}, we manually verify all the cases in Android 6 and find out 112 true positives of 133 possible vulnerabilities (see section~\ref{sec:tool_accuracy}). Unfortunately, we cannot thoroughly check all the possible vulnerabilities in various versions due to Android's large codebases, which is also a limitation in similar works such as Invettos~\cite{invetter} and Kratos~\cite{ss_attack_2}. Not all the detected vulnerabilities can be exploited, as some of them need special parameters that can not be constructed by third-party apps or need system-level permissions.

To further evaluate the effectiveness of detecting vulnerabilities, we construct POC apps to exploit these APIs and find out more than 32 exploitable cases. We reported some of the vulnerabilities to Google in 2017. Several countermeasures are taken since Android 8 includes fixing the bug case by case, adding standard permission checks, and using a blacklist to disallow directly invoking all the IPC interfaces. These countermeasures can make some of the previous exploitable bugs now are not vulnerable anymore. We run the exploitable POC in different Android versions and find out many of them are still vulnerable in newer Android versions (see Table~\ref{tab:bug-table}). Note that, here, we only discuss 23 bugs in section~\ref{sec:fixes} as the 9 IPC flood bugs have been fully discussed in~\cite{jni_exhuast}. 

\begin{table}[ht]
\centering
\caption{The number of classes, service helper classes, service IPC methods, and possible vulnerabilities report by ServiceAudit in different Android images.}
\begin{tabular}{|l|l|l|l|l|}
\hline \hline
\begin{tabular}[c]{@{}l@{}}Android\\ Version\end{tabular} & \#Class & \begin{tabular}[c]{@{}l@{}}\#Service \\ Helper\end{tabular} & \begin{tabular}[c]{@{}l@{}}\#Service \\ IPC API\end{tabular} & \begin{tabular}[c]{@{}l@{}}\#Possible\\ Vulnerabilities\end{tabular} \\ \hline \hline
6.1 (API 23) & 11577 & 178 & 1826 & 133 \\ \hline
7.1 (API 25) & 14705 & 173 & 1729 & 109 \\ \hline
8.0 (API 26) & 19854 & 218 & 2270 &  85 \\ \hline
9.0 (API 28) & 24166 & 268 & 2611 & 68 \\ \hline
10.0 (API 29) & 24166 & 285 & 2639 & 31 \\ \hline
11.0 (API 30) & 27582 & 317 & 2971 & 27 \\ \hline
12.0 (API 31) & 30856 & 358 & 3210 & 25 \\ \hline \hline
\end{tabular}

\label{tab:my-res}
\end{table}


\subsection{Tool Accuracy and Effectiveness}
\label{sec:tool_accuracy}

\noindent{\bf Accuracy of static analysis.} 
{We calculate true positives (TP) and false positives (FP) to evaluate the accuracy of our tool. Unfortunately, it is impossible to evaluate the false negatives as there are still zero-day bugs uncovered, and we cannot get the ground truth of all the vulnerabilities. We manually examine the source code of each system service methods and service helper methods to find out if the enforcement of service helper methods can be bypassed, e.g., for fake identify vulnerabilities, we verified whether the vulnerable API accept user identities such as userId and package name as parameters and do not verify these identities in system service API. If the protection can be bypassed, we count it as a true positive, and otherwise, we treat it as a false positive. And for the TP cases, we also construct test apps to try exploiting these vulnerable APIs. In all the 133 suspicious vulnerabilities, we find 112 true positives and more than 25 exploitable serious vulnerabilities. Comparing to other existing tools, our tools can detect more vulnerabilities with higher accuracy. The detailed results of each vulnerability categorization are shown in ~\ref{sec:results}.  }

\noindent{\bf Tool Effectiveness.} 
We further analyze the results of static analysis and try to understand why the false positives produce and the effect of exploiting the true positives. We find that most of the false-positive cases occurred due to the inaccuracy of point-to analysis and data-flow analysis and failed to identify the enforcements correctly, e.g., for IPC method \texttt{WallpaperService.engineShown()}, we mistakenly recognizing user identity checkings in its calling path. In fact, these checks are performed to protect another API. And another reason is that we cannot identify all the security enforcements in system services as there are too many ways to check the app's privilege. For example, for the IPC method \texttt{KeyStore.getState()}, we succeed in finding \textit{Uid} checking in the service helper method but failed to identify the security enforcement in system service API as it checks app permission via reading the authorization information from the ContentProvider.

{
\setlength\extrarowheight{3pt}
\begin{table}[ht]
\tinyv 
\caption{The results of our previous work and our current work compares to existing works.}
\centering
\begin{tabular}{ | c | c | c |c|c|}
\hline\hline

{\textbf{Work}} & {\textbf{Total Found}} & {\textbf{TP}} & {\textbf{Accuracy}} & {\textbf{Exploitable}} \\ [0.5ex] 
\hline\hline
Previous    & 143 & 32  & 22.4\% & 22  \\  \hline
Ours      	& 133 & 112 & 84.2\% & 32  \\ \hline
Kratos  	& 73  & 58  & 79.5\% & 8  \\ \hline
Invetter    & 103 & 86  & 83.5\% & 20  \\  \hline \hline
\end{tabular}
\label{tab:comparing}
\end{table}
}

{Not all the vulnerabilities can be exploited and cause serious security problems, such as \texttt{Toast.cancelToast()} can cancel other app's Toast message via fake-id attack by using other app's package name. Some vulnerabilities are hard to be exploited as they need special parameters and special logic states. Without static analysis, it is infeasible to verify all the IPC interfaces as there are a large number of methods, and many of them are difficult to construct test code. Our static analysis tool is effective in reporting the potential vulnerabilities with detail vulnerable information and reduce the manually verified methods number to less than 7\%. We find out at least 25 serious exploitable cases in all the reported 133 methods, and these issues are confirmed by Google. }

\noindent{\bf Tool Performance.} \texttt{ServiceAudit} needs about 4 to 8 minutes to finish the analysis of the whole \textit{farmework.jar}. It is much faster than our previous approach which requires several hours. However, if the Soot \textit{spart} point-to~\cite{soot} analysis is enable, our analysis needs more than one hour. Rather than using Soot's builtin call graph and point-to analysis, we propose partial call graph and point-to analysis that only focus on the service helper methods and the service IPC APIs and significantly reduce the analyzed classes. 

\noindent{\bf Compared with existing approaches.} 
Compared with Kratos~\cite{ss_attack_2} and Invetter~\cite{invetter}, our tool can discover more vulnerabilities with a higher accuracy shown in Table~\ref{tab:comparing}. The reports of Kratos are centered on fake identity vulnerabilities, and Invetter focuses on input related vulnerabilities such as fake identity and illegal parameters. Note that Kratos and Invetter support multiple Android versions, so we choose the results of Android versions, which are closest to the Android 6.0.1\_r1 used in our paper, i.e., Kratos's result for Android M preview (Android 6.0.0) and Invetter's result for AOSP (6.0). 
Among these works, Invetter also can checks if the service APIs have validated the caller's identity or permissions. For the vulnerability types which Invetter works on (input related), our tool can also detect most of the vulnerabilities given in their Table-4, such as \texttt{Accessibility Manager Service}, \texttt{Input Manager Service}, \texttt{Network Management Service}, \texttt{Audio Service}. However, our tool do not support the detection of vulnerabilities which depend on the parameter name, e.g., \texttt{CNEService} and \texttt{Atfwd} service . Due to the different working scopes, our tool can detect more types of exploitable vulnerabilities than Invetter such as the status checks bypass, and IPC flood.

\begin{figure}[t]
\centering
\includegraphics[width=0.48\textwidth]{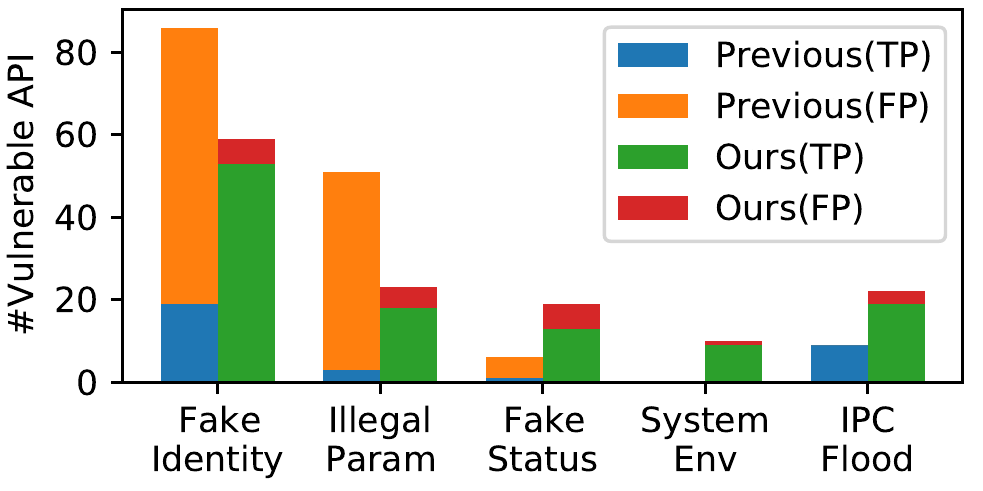}
\caption{The results of our current work and Service Helper~\cite{Gu16}. The number of true positives(TP) and false positives(FP) are concluded by manually verifying. Note that, the previous static tool does not support IPC flooding detecting so the false positives are not given. And system environment bypass is also new vulnerabilities discovered by the current tool.}
\label{fig:results}
\vspace{-10px}
\end{figure}
Compared to our previous approach~\cite{Gu16}, our overall true positives  (shown in Fig~\ref{fig:results}) have improved from 22.4\% to 84.2\% (reduced around 81\% false positives) on Android 6, which significantly reduced the overhead of manual confirmation. The aforementioned three improvements have greatly contributed to the better precisions. First, our new call graphs have correctly resolved virtual calls and thus can find the real service implementation classes while our previous approach may mistakenly use the parent classes as the service implementation, for instance, the BaseIDevicePolicy Manager which is an abstract class do not have any security enforcements, and thus produces lots of FP. Second, we use NLP approach (FPGrowth~\cite{fpgrowth}) to mine more security enforcements in system service sides that are not considered by our previous approach and reduce the FP of the fake identity vulnerabilities. Third, we use dataflow analysis to track if the parameters checked by service helpers are actually passed as IPC arguments to the system service and thus reduce the FP of the illegal parameter vulnerabilities, while our previous approach only checks the existence of some Exceptions. 



\subsection{Vulnerabilities Summary}
\label{sec:results}
We further study each vulnerability type and try to check if these vulnerabilities are exploitable. We find 178 service helper classes out of 11577 classes in Android 6.1. Among these service helpers, we find 133 potential vulnerabilities: 59 service helpers deliver the caller's identity to the service, which may suffer fake identity attacks. Also, we observe that 51 methods only validate input parameters in service helper methods and may suffer Illegal parameters attack, 19 methods that verify caller's status and may suffer the fake status attack, and 22 service methods that handle duplicated requests without protecting, and ten methods evaluating system environments.

After manually verifying these potential vulnerabilities in Android 6.1, we find 112 true positives. From these true positives, we further construct POC apps to find out the vulnerabilities that can be exploited and lead to severe problems. Note that not every case can be exploited as some may need special arguments that can not be constructed by the user, and some may do not have any harm even the checks are bypassed. we capture 32 vulnerabilities that lead to serious security issues, e.g., privilege escalation, {data leakage}, user interactions bypass, or Android system soft reboot, shown in Table~\ref{tab:bug-table}. All these vulnerabilities have been confirmed by the Android Security Team and assigned with different Android Bug IDs. For the left vulnerabilities, some are fixed in the later Android, and some can be exploited but do not cause serious security problems. We do not submit them to Google as AOSP 6 is out of date. We also revisit these bugs in newer Android versions and find out that these problems still exist. We describe the observed vulnerabilities in the following. 

\noindent{\bf Vulnerabilities Caused by Passing Illegal Parameter.}
We identify 227 service helper methods containing parameter checking and exception throw \texttt{stmt}. In fact, only a few of these checks are used to protect the system service API and we perform inter-procedure dataflow analysis to verify if the input parameters of IPC methods are really checked. Then we apply permission filtering to exclude the IPC methods which require special permissions and difficult to exploit. Most of these methods are excluded after the two approaches and they help to reduce to false positive rate from 94\%(48/51) to less than 20\%(5/23).
And we find 23 method pairs are inconsistent in validating the input parameters and 18 methods of them are true positives could be possibly exploited. After manual verification, three methods are confirmed to be vulnerable with serious problems, i.e., they can be exploited to crash their services. These vulnerabilities are in \texttt{MediaBrowserService} and \texttt{HealthService}, which are incurred by failure in handling exceptions.


\noindent {\bf Vulnerabilities Caused by Passing Fake Identities.}
In our previous approach, we find 86 inconsistent identity verification, however, it relies on the manually set sensitive API list to identity security checking methods and most of the system service API use custom checking methods which produces lots of false positives. In our current approach, after performing associate rules mining, we identity more security checking methods in both service helper and system service. We have discovered 59 inconsistent methods and most of them are new cases missing by our previous tool. The false positive rate has reduced to less than 10\%.
In all these reported inconsistent methods, we find 53 true positives that service methods receive callers' identities from the corresponding service helpers but fail to verify the authenticity of the received identities as the service helpers do.  We manually verify whether all of them can be exploited. Our verification shows that 10 out of the 53 inconsistent method pairs can be used to launch real-world attacks. The rest methods are not vulnerable to fake identity attacks because those fake identities do not incur security issues, such as \texttt{BackupManager.dataChanged()}. 

Among the exploitable vulnerabilities, one case is caused by the abuse of \texttt{enqueueToast()} in the notification service, which can lead to system reboot.
Malicious apps will be falsely treated as a system app via passing a fake package name ``android'' and thus exhausts the system resources. 
The other nine vulnerabilities are in the \texttt{notification} service and the \texttt{fingerprint} service, respectively, which will result in privilege escalation.
A real-world attack to \texttt{fingerprint} service is illustrated in Section~\ref{sec:hazards}. 

\noindent {\bf Vulnerabilities Caused by Invoking IPC with Fake Status.}
We find that 19 service helper methods check the caller's status, and 13 of them are true positives. One of them lacks the validation of the caller's status in its corresponding service method, i.e., \texttt{NfcService}. An attacker can easily bypass user interaction to access function without user initiation or user permission.  We show this case in Section~\ref{sec:hazards} 
where an app can directly retrieve NFC data when it is in the background. And some other APIs can also be exploited but do not cause serious problems, such as AccessibilityManager's status checking can also be bypassed which may be exploited by attackers and interrupt other apps' accessibility services.


\noindent{\bf Vulnerabilities Caused by Bypassing System Environment Evaluation.} 
We find 10 methods that evaluate system environment in different service helpers and identify 9 possible vulnerabilities.  
Most service helpers only handle encapsulating the functions of system services. Here, we only identify one vulnerable method, i.e., \texttt{WallpaperManager.isWallpaperSupported()}, in evaluating system setup, which incurs data leakage used in \texttt{WallpaperManagerService.getWallpaper()}.

\noindent{\bf Vulnerabilities Caused by IPC Flooding.}
We identify 22 methods in service helpers that handle duplicated requests and can be bypassed. 
These helper methods firstly check whether the current request is a duplicated one. 
After receiving duplicated requests, the methods either process the requests locally (not deliver them to the services) or constrain the number of the requests that can be delivered to the services. 
However, these methods can be easily bypassed by directly using the methods in the corresponding system services.
A malicious app can directly invoke a system service via IPC without any restriction. We find that a large number of IPC calls would lead to Android resource exhaustion, which can further incur system reboot. And some vulnerabilities do not cause to reboot but can make the system frozen, such as \texttt{AudioManager.setStreamVolume()}. If these APIs are called too many times the floating audio adjusting, window will make the system frozen and have no response.


{
\def\Y{$\bullet$}
\def\C{\ding{52}}
\def\X{\ding{55}}

\begin{table*}[ht]
\centering
\caption{The exploitable vulnerabilities (exclude 9 IPC flood cases) and their impacts and mitigate effectiveness in various Android versions. Google has taken several countermeasures includes, put API in Blacklist on Android 11 (B), fix the bug (F), and API has been removed or changed (C).}
\begin{tabular}{|l|l|cccc|l|c|}
\hline\hline
\multirow{2}{*}{\textbf{Service Name}} & \multicolumn{1}{c|}{\multirow{2}{*}{\textbf{Vulnerable API}}} & \multicolumn{4}{c|}{\textbf{Affected Frameworks}} & \multicolumn{1}{c|}{\multirow{2}{*}{\textbf{Detail}}} & \multirow{2}{*}{\textbf{Mitigate}} \\ \cline{3-6}
 & \multicolumn{1}{c|}{} & \textbf{6.1} & \textbf{8.0} & \textbf{10.0} & \textbf{11.0} & \multicolumn{1}{c|}{} &  \\ \hline\hline
TvInputManagerService & selectTrack & \Y &  &  &  & Select TV track with the fake uid & C \\ \hline
\multirow{5}{*}{AudioService} & setStreamVolume & \Y & \Y & \Y &  & Set volume without permission & F \\ \cline{2-8} 
 & setRingerModeExternal & \Y & \Y & \Y &  & Modify volume settings & B + F \\ \cline{2-8} 
 & adjustStreamVolume & \Y & \Y & \Y &  & Adjust volume & B + F \\ \cline{2-8} 
 & setBluetoothA2dpOn & \Y & \Y &  &  & Disable A2DP audio & B + F \\ \cline{2-8} 
 & setBluetoothScoOn & \Y & \Y &  &  & Disable Bluetooth SCO headset & B \\ \hline
\multirow{5}{*}{NotificationManagerService} & enqueueToast & \Y & \Y &  &  & DoS attack with too many toast & B + F \\ \cline{2-8} 
 & cancelToast & \Y & \Y & \Y &  & Cancel system notification & B + F \\ \cline{2-8} 
 & setNotificationPolicy & \Y &  &  &  & Change notification policy & B + F \\ \cline{2-8} 
 & getNotificationPolicy & \Y &  &  &  & Get notification policy & B + F \\ \cline{2-8} 
 & setInterruptionFilter & \Y &  &  &  & Change notification filter & B + F \\ \hline
IInputMethodSession & displayCompletions & \Y & \Y & \Y &  & Close the input method panel & B + F \\ \hline
DisplayManagerService & releaseVirtualDisplay & \Y & \Y &  &  & Relase the virtual display & B + F \\ \hline
\multirow{5}{*}{FingerprintService} & authenticate & \Y & \Y &  &  & Start fingerprint authenticate & B + F \\ \cline{2-8} 
 & cancelAuthentication & \Y & \Y &  &  & Stop fingerprint authenticate &  B + F  \\ \cline{2-8} 
 & getEnrolledFingerprints & \Y & \Y &  &  & Get fingerprint status &  B + F  \\ \cline{2-8} 
 & hasEnrolledFingerprints & \Y & \Y &  &  & Get fingerprint status &  B + F  \\ \cline{2-8} 
 & isHardwareDetected & \Y & \Y &  &  & Get fingerprint status &  B + F  \\ \hline
\multirow{2}{*}{MediaBrowserService} & addSubscription & \Y & \Y & \Y &  & DoS with illegal parameter &  B + F  \\ \cline{2-8} 
 & removeSubscription & \Y & \Y & \Y &  & DoS with illegal parameter &  B + F  \\ \hline
HealthService & registerAppConfiguration & \Y &  &  &  & DoS with illegal parameter & F \\ \hline
NfcAdapter & enableForegroundDispatch & \Y & \Y &  &  & Change NFC settings &  B + F  \\ \hline
WallpaperManager & getWallpaper & \Y & \Y &  &  & Read wallpaper & F\\ \hline

ShortcutService & requestPinShortcut & - & - & \Y & \Y &  Create App Shortcut & F \\ \hline\hline
\end{tabular}
\label{tab:bug-table}
\end{table*}

\undef\Y
}

\subsection{The Effectiveness of Android Countermeasures}
\label{sec:fixes}
To answer whether, there are still exploitable vulnerabilities in hidden APIs. We applied our tool to the newest Android 11, 12 and found more than 25 vulnerabilities. To further understand the hazards of exploitation, we manually verified all these APIs on Android emulators. To estimate the effectiveness of Google’s countermeasures, we study how previous vulnerable APIs are fixed in newest Android 12 and how many vulnerabilities are still exploitable in Android 12. Our findings are as follows.

\noindent \textbf{Bugs are Eliminated or Emerge due to Version Changes. } During the development of the Android framework, some existing APIs may have been refactored or removed in new versions, such as the \texttt{selectTrack()} method from \texttt{TvInputManagerService}. We find some vulnerable APIs have been changed and can not be exploited anymore.  It is interesting to find that the newly added APIs may also be vulnerable, such as the \texttt{requestPinShortcut} of the \texttt{ShortcutService} is introduced since Android 9 and it is not be fixed since Android 12. 

\noindent \textbf{Most of the Bugs are Fixed.} The vulnerabilities are gradually fixed after Android 8 by adding extra security enforcements in the callee system service APIs’ side. We revisit all the serious vulnerabilities in previous versions, such as, set audio without permission, or stealing the Clipboard. We find that Google has added security checks to almost every public API in \texttt{AudioService}. In \texttt{ClipboardService}, they have now validated the permission of the real callers of the pending Intent via \texttt{getIntendingUid}.


\noindent \textbf{The Remained Vulnerable APIs Seems to Be Secure.} We have identified 27 vulnerabilities in Android 11 and two of them are eliminated in Android 12. Almost all of these vulnerabilities seem to have no harms to be exploited since most of them are APIs to get status, such as \texttt{getStreamVolume}, \texttt{hasSystemFeature}. For the \texttt{StorageStatsService}, the \texttt{getCacheBytes} APIs requires the permission \texttt{PACKAGE\_USAGE\_STATS} while a similar API \texttt{getTotalBytes} can be invoked without permission but with  a comment of \textit{"No permissions required"}. Although some other APIs in \texttt{StorageStatsService} are not marked with no permission required, we do not regard them as dangerous since they can only be used to obtain status, e.g., \texttt{isReservedSupported}.  For the vulnerabilities that can bypass the system environment evaluating, we also failed to find any dangerous one, e.g., \texttt{startWatchingRoutes} seems can only obtain the status of the \texttt{AudioRoutesInfo}. In Android 11, we only find the \texttt{requestPinShortcut} can incur security problems by arbitrarily creating Icon shortcuts~\cite{shortcut_11} for other apps. Fortunately, it is fixed~\cite{shortcut_12} in Android 12.   

\begin{figure}[t]
\centering
\includegraphics[width=0.48\textwidth]{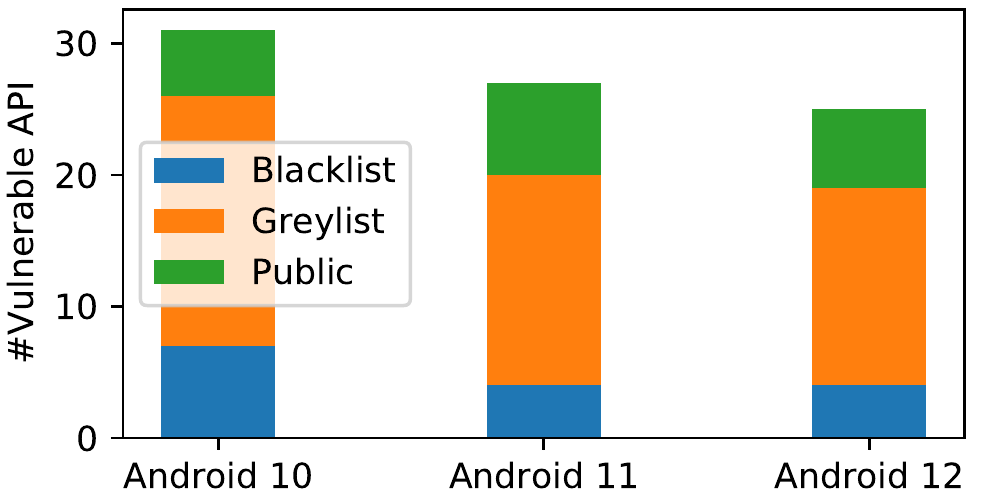}
\caption{The numbers of vulnerable APIs with different restriction policy.}
\label{fig:blacklist}
\vspace{-10px}
\end{figure}

\noindent \textbf{Blacklist or Greylist does no Affect the Hidden API Security.} We statistic how many vulnerabilities APIs are in blacklist or greylist. As shown in Fig~\ref{fig:blacklist}, most of these APIs are in blacklist and greylist. It reveals that the developers need to evade the no-SDK APIs restriction to exploit the vulnerable APIs since the blacklist APIs can not be used and the greylist APIs can only be used before Android 9. Considering, even if the attackers evade the hidden APIs restriction via the methods in Section~\ref{sec:evade}, the exploited APIs have no security risk. 

\section{Related Work}
Android vulnerabilities have been extensively studied in the literature, e.g., private data leakage~\cite{ app_1,app_3,app_4,app_6,app_7}, privilege escalation~\cite{ p_e_1,p_e_2,p_e_3,app_11,p_e_4}, and component hijacking~\cite{app_8,app_9,app_10,app_12}.  
In this paper, we only summarize and compare with the existing studies closely related to the service helper bypass vulnerabilities.

\subsection{Android Static Analysis}
Static analysis is one of the most effective techniques to analyze the vulnerabilities in both Android systems and apps.
There have been various studies on malware detection by leveraging static analysis~\cite{Enck:2009:LMP:1653662.1653691,fuchs2009scandroid,Grace:2012:RSA:2307636.2307663}, library security~\cite{Grace:2012:UEA:2185448.2185464}, repackaging detection~\cite{app_13,Hanna:2012:JSS:2481803.2481809}, component security~\cite{app_12}, system service security~\cite{ss_attack_1,ss_attack_2}, and permission specification~\cite{pscout}. 
Several static analysis tools~\cite{soot,desnos2013androguard,androbugs} also have been developed to solve different problems.
There are two tools~\cite{ss_attack_1,ss_attack_2} closely related to our work. Our study focuses on the security breaches incurred by bypassing service helpers, which cannot be discovered by them. 
Moreover, since the static analysis cannot accurately reflect the precise situations in runtime, the analysis results are accurate. In our paper, we automatically discover the vulnerabilities and verify the found vulnerabilities by constructing real-world attacks.

\subsection{Vulnerability Detection in System Service}
Invetter~\cite{invetter} and Kratos~\cite{ss_attack_2} are also studying the vulnerabilities due to inconsistent security policy enforcement of system service. Invetter studies the sensitive inputs and verifies if the protection of these sensitive inputs is consistent between system services methods and service helpers methods. Invetter uses a machine learning approach to identify new sensitive inputs via learning from existing parameter names to detect if these parameters are well protected. Different from their approach, we discover insufficient security checks of the system APIs (IPC.stub) by comparing their security enforcements to the service helpers (IPC.proxy clients) to mine the inconsistency. By mining new security mechanisms (e.g., permission checks, status checks, sanity checks) instead of sensitive input, our approach support more types of vulnerabilities other than the input related vulnerabilities. Invetter has reported 9 exploitable services in AOSP 6.0 and 11 vulnerabilities in other third-party ROMs. In all the 9 vulnerabilities of AOSP 6.0, 8 cases are also reported by our tool and only the \texttt{CNEService} is missed. Kratos finds out that multiple execution paths are leading to the same system service function with inconsistent privilege requirements. Thus, malicious apps can easily escalate their privileges or even perform DoS attacks by redirecting their requests to the paths with less enforced permissions. Different from these studies only on the system services, our work focuses on investigating the impacts of bypassing service helpers and studying the security mechanisms in both services and service helpers. And we perform a systematic and comprehensive study about several types of vulnerabilities due to inconsistent security enforcements of system services.


\subsection{Android System Service Security}
The system services security have been studied in various aspects~\cite{ss_attack_1,ss_attack_2,ss_attack_3, ipl_service, mes_service}. Lei et al.,~\cite{ipl_service} found a new kind a service hijack attack due to implicit services and measure~\cite{mes_service} the vulnerable Apps of different time periods in Google Play Market.
Huang et al.,~\cite{ss_attack_1} discovered a design flaw in the concurrency control of Android system services. They found that Android system services often used the lock mechanism to protect critical sections or synchronized methods. However, if an application took a lock for a long time and other system services sharing the same lock would freeze, then the watchdog thread would force Android to reboot. 
Another related work~\cite{ss_attack_3} examines the input validation in system services using fuzzing. They have identified several DoS attacks due to the lack of proper input validation in system services.
Our work has also examined the input validation of system services. We also found that the App identity collected by the service helper and the parameters prepared by the developer are passed to the service as its input which may suffer attacks.
We present an effective approach to identify the vulnerabilities. As shown in Section~\ref{sec:results},  we systemically reveal more vulnerabilities~\cite{jni_exhuast} other than the parameter validation that were missed by previous approaches. 
The existing studies~\cite{ss_attack_3, invetter} are also unable to discover such vulnerabilities since the vulnerabilities can only be exploited by constructing special parameters. For example, the \texttt{FingerprintService} service can be exploited if the input parameter is set to be the package name of \texttt{KeyguardService} (see Section~\ref{sec:hazards}).
However, it is still challenging to construct the parameters by fuzzing to effectively find this vulnerability. 

\section{Discussion}

\subsection{Lessons Learned}
Our previous study reveals a large number of vulnerable hidden service APIs (Section~\ref{sec:Findings}), which can lead to serious security issues, e.g., privilege escalation, user interaction bypass, service crash, and Android system soft reboot. Since Google cannot prevent the attacker from exploiting the hidden service APIs, the only way to mitigate this problem is to eliminate all the bugs by manually adding enough security checks in all the system service APIs. Considering the huge numbers of hidden service APIs, it is critical to use static analysis tools to report such problems.

\subsection{Limitations}
\noindent \textbf{Detection Accuracy.} Although we capture a large number of vulnerabilities caused by the bypass of service helpers, we have to admit that there may exist more such vulnerabilities to be uncovered.  
The main reason is that some system services use native code with JNI though it is a small portion of the service code on the Android system~\cite{ss_attack_2}. 
We do not study the native code in our analysis yet, which will result in false negatives. 
We leave the study of vulnerable system services with JNI native code incurred by bypassing service helpers as future work. Note that, we could use fuzzing to identify vulnerabilities. However, without the information in the system service helpers, the efficiency of fuzzing is low. In the future, we can try to leverage fuzzing to improve the efficiency of identifying the vulnerabilities. 

\noindent \textbf{Manual Work.} Our approach is mostly performed automatically. However, manual verification is inevitable since we need to confirm identified vulnerabilities by constructing PoCs, which cannot be automatic. For example, we find that passing fake IDs to some system services will not incur any security problems in the status obtain functions.  
Normally, manual work is required in verifying whether the identified vulnerabilities can indeed be exploited to launch real-world attacks. 
The procedure is necessary since we can systematically investigate the impacts of the vulnerabilities by verifying if they can be exploited.  
\section{Conclusion}
We systematically study the non-SDK service API security and focus on the vulnerabilities of bypassing service helpers and corresponding countermeasures in various Android versions.  In order to discover such vulnerabilities and demonstrate the impacts of the vulnerabilities, we develop a methodology that automatically analyzes and compares the security mechanisms in system services with the associated service helpers in various AOSP. 
We study the effectiveness of Google's countermeasures in different Android versions and check if the vulnerable APIs are still exploitable in newest Android 12.  In Android 6, we discovered 32 serious vulnerabilities that can be used to launch real-world attacks, while in Android 12, only 25 APIs are reported as vulnerable but have no harm to be exploited. The experimental results reveal that Google has successfully addressed the vulnerabilities in hidden APIs due to inconsistent security enforcements.



\bibliographystyle{plain}
\bibliography{main}

\end{document}